\documentclass[]{JHEP3}

\usepackage{epsfig}
\usepackage{amsmath,amssymb}
\usepackage{citesort}

\newcommand{\nn}{\nonumber}
\newcommand{\sig}{\sigma}
\newcommand{\lam}{\lambda}
\newcommand{\eps}{\epsilon}    
\newcommand{\alp}{\alpha}
\newcommand{\bet}{\beta}

\newcommand{\M}{{\cal M}}  
\newcommand{\J}{{\cal J}}  
\newcommand{\Mh}{\hat{\cal M}}
\newcommand{\Jh}{\hat{\cal J}}
\renewcommand{\P}{{\cal P}}  
\newcommand{\D}{{\cal D}}

\title{Jet angular correlation in vector-boson fusion processes 
       at hadron colliders}

\author{Kaoru Hagiwara$^a$, Qiang Li$^b$ and Kentarou Mawatari$^{c,d}$\\
        $^a$KEK Theory Division and Sokendai,
        Tsukuba 305-0801, Japan\\
        $^b$Institut f\"ur Theoretische Physik, Universit\"at Karlsruhe,
        Postfach 6980, D-76128 Karlsruhe, Germany\\
	$^c$Institut f\"ur Theoretische Physik, Universit\"at
        Heidelberg, Philosophenweg 16, D-69120 Heidelberg, Germany\\
 	$^d$School of Physics, Korea Institute for Advanced Study,
        Seoul 130-722, Korea\\
        E-mail: \email{qliphy@particle.uni-karlsruhe.de,
                       k.mawatari@thphys.uni-heidelberg.de}}

\abstract{
Higgs boson and massive-graviton productions in association with two
jets via vector-boson fusion (VBF) processes and their decays into a
vector-boson pair at hadron colliders are studied.  
They include scalar and tensor boson production processes via weak-boson
fusion in quark-quark collisions, gluon fusion in quark-quark ($qq$),
quark-gluon ($qg$) and gluon-gluon ($gg$) collisions, as well as their
decays into a pair of weak bosons or virtual gluons which subsequently
decay into $\ell\bar\ell$, $q\bar q$ or $gg$. 
We give the helicity amplitudes explicitly for all the VBF subprocesses, 
and show that the VBF amplitudes dominate the exact matrix elements not
only for the weak-boson fusion processes but also for all the gluon
fusion processes when appropriate selection cuts
are applied, such as a large rapidity separation between two jets and a 
slicing cut for the transverse momenta of the jets.
We also show that our off-shell vector-boson current amplitudes reduce
to the standard quark and gluon splitting amplitudes with appropriate
gluon-polarization phases in the collinear limit.
Nontrivial azimuthal angle correlations of the jets in the
production and in the decay of massive spin-0 and -2 bosons are
manifestly expressed as the quantum interference among
different helicity states of the intermediate vector-bosons.
Those correlations reflect the spin and the $CP$ nature of the Higgs
bosons and the massive gravitons.}   

\preprint{KEK-TH-1219\\ KA-TP-08-03\\ SFB/CPP-08-03\\ HD-THEP-09-6\\
          KIAS-P08016\\[1mm]}

\begin{document}

\section{Introduction}

Angular correlation of the two accompanying jets in Higgs boson 
productions at the CERN Large Hadron Collider (LHC) has been known 
as a potential tool to study its spin and $CP$ nature, in the weak-boson
fusion (WBF) $qq\to qqH$ processes~\cite{Plehn:2001nj}, and in the gluon
fusion (GF) plus dijet production 
processes~\cite{DelDuca:2001eu,Hankele:2006ja,Klamke:2007cu}, 
$qq\to qqH$, $qg\to qgH$ and $gg\to ggH$. In these reactions,  
the tensor structure of the Higgs coupling to weak bosons 
or gluons gives rise to the azimuthal angle correlation of the tagging 
jets; the WBF processes
give flat $\Delta\phi_{jj}$ distribution, while the GF processes
produce a distinct dip around $\Delta\phi_{jj}=\pi/2$.  
On the other hand, in the case of a $CP$-odd Higgs boson, 
the azimuthal distribution is
strongly enhanced around $\Delta\phi_{jj}=\pi/2$ in both the WBF and GF
processes~\cite{Plehn:2001nj,Hankele:2006ja,Klamke:2007cu}.%
\footnote{Azimuthal correlations in diffractive processes have also
been discussed in 
refs.~\cite{Arens:1996xw,Close:1999is,Kaidalov:2003fw}.} 

So far, many studies on the azimuthal correlations in the Higgs + 2-jet 
events have been performed with higher-order QCD and electroweak
corrections~\cite{Figy:2004pt,Campbell:2006xx,Hankele:2006ma,Ciccolini:2007jr,Andersen:2007mp,Andersen:2008gc} 
including parton-shower effects~\cite{Odagiri:2002nd,DelDuca:2006hk}.
The present consensus seems to be that the azimuthal angle correlations
predicted in the leading order may survive even after higher-order
corrections are applied~\cite{DelDuca:2006hk,Andersen:2008gc}.  
It must be pointed out here that, due to the complicated matrix elements,
it is not completely clear why the tensor structure of the couplings
leads to such distinct azimuthal distributions, even though naive
explanations have been presented~\cite{Plehn:2001nj,DelDuca:2001eu}. 
Since, in general, azimuthal angle dependence should be understood in
terms of the quantum interference phases of the amplitudes with
spin-full particles propagating along the polar axis, it may be
valuable to reformulate the amplitudes for the Higgs production with two
jets in such a way that their phases are shown explicitly. 

As another interest to study the azimuthal angle correlation of the
jets at the LHC, we attempt to apply it to other heavy particle
productions. Here, we especially focus on massive-graviton productions
in the localized gravity model of Randall and Sundrum 
(RS)~\cite{Randall:1999ee}, which has drawn a lot of attention in
recent years because it brings a new solution to the hierarchy problem
through an exponentially suppressed warp factor in a 5-dimensional
non-factorizable geometry. 
Several phenomenological studies have been made on the
Drell-Yan process for RS graviton resonances for its discovery
and the determination of its spin-2 
nature~\cite{Davoudiasl:1999jd,Allanach:2000nr,Allanach:2002gn,Traczyk:2002jh,Osland:2003fn,Cousins:2005pq,Mathews:2005zs},
including direct searches at the Tevatron~\cite{Abazov:2005pi}, 
as well as the graviton + 1-jet productions~\cite{Murayama:2009jz}.
Meanwhile massive-graviton productions in association with two jets may
also have a great potential to scrutinize its properties as in the Higgs
boson case. 
We note that the graviton + 2-jet productions in the large extra
dimensions model~\cite{ArkaniHamed:1998rs} have recently been studied in
ref.~\cite{Hagiwara:2008iv}.\\ 

In this article, more generally, we study productions of a heavy
color-singlet particle ($X$) in association with two jets via
vector-boson fusion (VBF) processes at hadron colliders, $pp\to jjX$,
which include WBF processes in quark-quark collisions and GF processes
in quark-quark, quark-gluon and gluon-gluon collisions.%
\footnote{Weak-boson fusion is sometimes referred to as VBF. In this
 paper, however, we refer to the fusion processes of all the standard model
 vector-bosons ($W,Z,\gamma,g$) as VBF, including WBF and GF processes.}
In particular, the reactions,
\begin{align}
 qq\to qqX,\quad qg\to qgX,\quad gg\to ggX,
\end{align}
are studied comprehensively as the leading-order subprocesses that lead
to $X$ + 2-jet events via VBF.    
In order to discern the phases of the amplitudes, we present the
helicity amplitudes explicitly for all the VBF subprocesses at the tree
level in terms of the specific kinematical variables, where the
colliding vector-bosons have momenta back-to-back along the polar axis. 
Although the VBF amplitudes are valid only when the virtuality of
the intermediate vector-bosons is smaller than their energies, as we
will see later, they can dominate the exact matrix elements when
appropriate selection cuts to the final states are applied, such as a large
rapidity separation between two tagging jets and a slicing cut for the 
transverse momenta of the jets. 
We also show that our off-shell vector-boson current amplitudes reduce
to the standard quark and gluon splitting amplitudes with appropriate
phases in the collinear limit.

As for the produced heavy particles, we study neutral $CP$-even and
$CP$-odd Higgs bosons and RS massive gravitons, and show that 
nontrivial azimuthal angle correlations of the jets in the
production of massive spin-0 and -2 bosons are
manifestly expressed as the quantum interference among
different helicity states of the intermediate vector-bosons.
Those correlation reflects the spin 
and the $CP$ nature of the Higgs bosons and the massive gravitons.
We do not consider massive spin-1 particles because the Landau-Yang theorem
forbids production of a color-singlet spin-1 particle in fusions of two
on-shell photons or gluons~\cite{Yang:1950rg}, and our approximation
fails when their virtuality is large enough to give significant amplitudes. 

Besides jet angular correlations in the production processes, the decay
distributions and correlations of heavy particles are also promising
tools to determine their properties, and extensive studies have been
made especially for the Higgs bosons, {\it e.g.} 
$H\to ZZ\to(\ell\bar\ell)(\ell\bar\ell)$~\cite{Dell'Aquila:1985ve,Skjold:1993jd,Hohlfeld:2001,Choi:2002jk,Buszello:2002uu}; 
see also review papers~\cite{Djouadi:2005gi,Rainwater:2007cp} and
references therein. 
The above decay process is related by crossing symmetry to the WBF Higgs
production process, $qq\to qqH$, and hence it may be useful to compare
the production correlations with the decay correlations.   
Therefore, we also consider $X$ decays into a pair of weak-bosons or
gluons which subsequently decay into $\ell\bar\ell$, $q\bar q$ or $gg$,
and present the helicity amplitudes and the azimuthal angle correlations
of the jets (and/or leptons), by comparing with those in the production
processes.\\

The article is organized as follows. 
In section~\ref{sec:formalism}, we introduce the formalism of the
helicity amplitudes and the density matrices for
the $X$ production with two jets via VBF and the $X$ decay into a
vector-boson pair. 
In section~\ref{sec:kinematics} we define kinematical variables
relevant to our analysis for the production and decay processes. 
In section~\ref{sec:helamp}, we present all the helicity amplitudes
explicitly for the off-shell vector-boson currents and the off-shell VBF
processes. We also discuss the relation between our 
current amplitudes and the standard parton splitting amplitudes.
In section~\ref{sec:az}, we demonstrate that the VBF amplitudes
dominate the exact matrix elements when appropriate selection cuts to 
the final states are applied, and then discuss 
azimuthal angle correlations of the jets in the Higgs boson and
massive-graviton productions.
We also consider the decay correlations of the heavy particles. 
Finally section~\ref{sec:summary} summarizes our findings.

We include three appendices. 
Appendix~\ref{sec:spin2} gives the wavefunction and the vertices 
for a spin-2 particle. In appendix~\ref{sec:dfunc} we show the
relation between wavefunctions and Wigner's $d$~functions.
Appendix~\ref{sec:gravitondecay} presents the angular
distributions for the massive-graviton decays, 
$G\to VV\to (f\bar f)(f\bar f)$.

\section{Helicity formalism}\label{sec:formalism}

In this section, we give the helicity amplitude formulae and the density
matrix formalism for heavy
particle ($X$) productions in association with two jets via VBF
processes, and also those for its decay into a (virtual) vector-boson
pair which subsequently decay into $\ell\bar\ell$, $q\bar q$ or $gg$.\\  

\FIGURE[t]{
 \epsfig{file=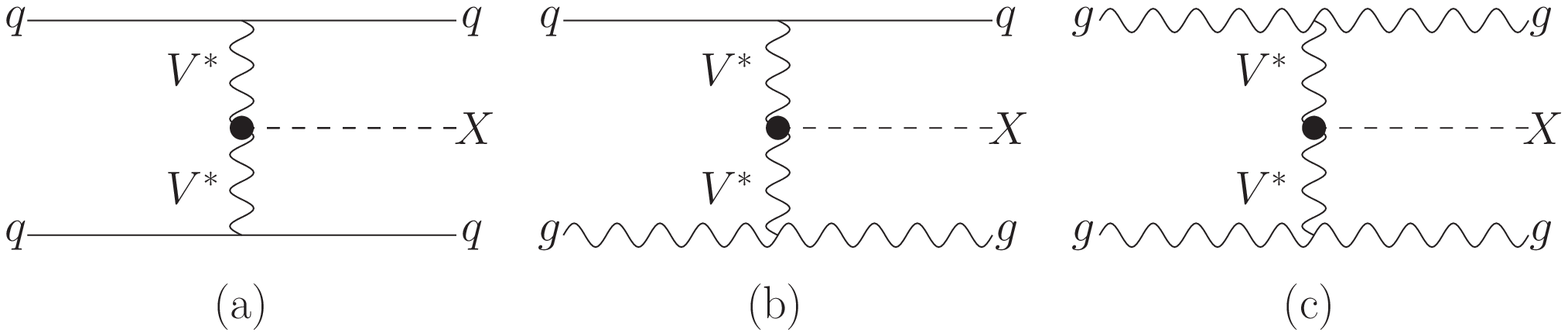,width=1\textwidth,clip}
 \caption{Feynman diagrams for the VBF subprocesses: (a) $qq\to qqX$,
 (b) $qg\to qgX$, and (c)~$gg\to ggX$.
 \label{fig:vbf}}}

$X$ + 2-jet productions via VBF at hadron colliders, $pp\to jjX$, can
proceed through the subprocesses: 
\begin{subequations}
\begin{alignat}{4}
 qq &\to qqV^*V^* &&\to qqX &\qquad&(V=W,\,Z,\,\gamma,\,g),
\label{qq_qqX}\\ 
 qg &\to qgV^*V^* &&\to qgX && (V=g),
\label{qg_qgX}\\
 gg &\to ggV^*V^* &&\to ggX && (V=g),
\label{gg_ggX}
\end{alignat}
\label{ff_ffx}%
\end{subequations}
where $V^*$ is a $t$-channel intermediate vector-boson and $q$ stands
for a quark or antiquark of any flavors. A representative  
Feynman diagram for each subprocess is shown in fig.~\ref{fig:vbf}(a),
(b), and (c), respectively, for the subprocess (\ref{qq_qqX}),
(\ref{qg_qgX}), and (\ref{gg_ggX}). 
Each subprocess receives contributions not only from the above VBF
diagram but also from all the other diagrams of the same order, in order
to make the gauge-invariant physical amplitudes. In this section,
however, we consider only the VBF diagram. 
As we shall see later in section~\ref{sec:az}, after applying
appropriate kinematical selection cuts, the VBF contribution can dominate
the exact matrix elements. 

Let us first define a common set of kinematical variables for the VBF
subprocesses (\ref{ff_ffx}) generically as 
\begin{align}
      a_1(k_1,\sig_1)+a_2(k_2,\sig_2) 
 &\to a_3(k_3,\sig_3)+a_4(k_4,\sig_4)
     +V_1^*(q_1,\lam_1)+V_2^*(q_2,\lam_2) \nn\\  
 &\to a_3(k_3,\sig_3)+a_4(k_4,\sig_4)+X(P,\lam),
\label{qq_qqx}
\end{align}
where $a_{1,\cdots,4}$ stand for quarks or gluons (or even leptons in
case of lepton-lepton or lepton-hadron collisions), and 
the four-momentum and the helicity of each particle are shown in
parentheses; see also fig.~\ref{fig:diagram}(a). The parton helicities
take the values $\sigma_{i}/2$ for quarks or antiquarks 
and $\sigma_{i}$ for gluons
with $\sigma_i=\pm 1$, while the helicities of the off-shell vector-bosons
take $\lambda_i=\pm 1,0$.

The helicity amplitudes for the VBF processes (\ref{ff_ffx}) can 
generally be expressed as
\begin{align}
 \M^{\lam}_{\sig_1\sig_3,\sig_2\sig_4} = \sum_{V_{1,2}}\, &
 J^{\mu'_1}_{V_1a_1a_3}(k_1,k_3;\sig_1,\sig_3)\,
 J^{\mu'_2}_{V_2a_2a_4}(k_2,k_4;\sig_2,\sig_4) \nn \\[-2mm] &\times    
 D^{V_1}_{\mu'_1\mu^{}_1}(q_1)\,D^{V_2}_{\mu'_2\mu^{}_2}(q_2)\,
 \Gamma^{\mu_1\mu_2}_{XV_1V_2}(q_1,q_2;\lam)^*,
\label{amp}
\end{align}
where $J^{\mu}_{V_1a_1a_{3}}$ and $J^{\mu}_{V_2a_2a_{4}}$ are 
the external fermion or gluon currents, and  
the vector-boson propagators are 
\begin{align}
 D^{V_i}_{\mu'\mu}(q_i)=
 \begin{cases}
   \Big(-g_{\mu'\mu}+\dfrac{{q_i}_{\mu'}{q_i}_{\mu}}{m_{V_i}^2}\Big)
   D_{V_i}(q_i^2)
   & \text{for}\ V_i=W,\,Z, \\[3mm]
   -g_{\mu'\mu}\,D_{V_i}(q_i^2)
   & \text{for}\ V_i=\gamma,\,g,
 \end{cases}  
\end{align}
with the propagator factor $D_{V}(q^2)=(q^2-m_V^2+im_V\Gamma_V)^{-1}$.
Note that we choose the unitary-gauge propagator for the massive 
vector-bosons and the Feynman-gauge one for the massless vector-bosons
($m_{V_i}=0$).
The $XVV$ vertex is expressed generically as 
$\Gamma^{\mu_1\mu_2}_{XV_1V_2}$, whose explicit forms are given in
section~\ref{sec:vbfamp}. 

\FIGURE[t]{
 \epsfig{file=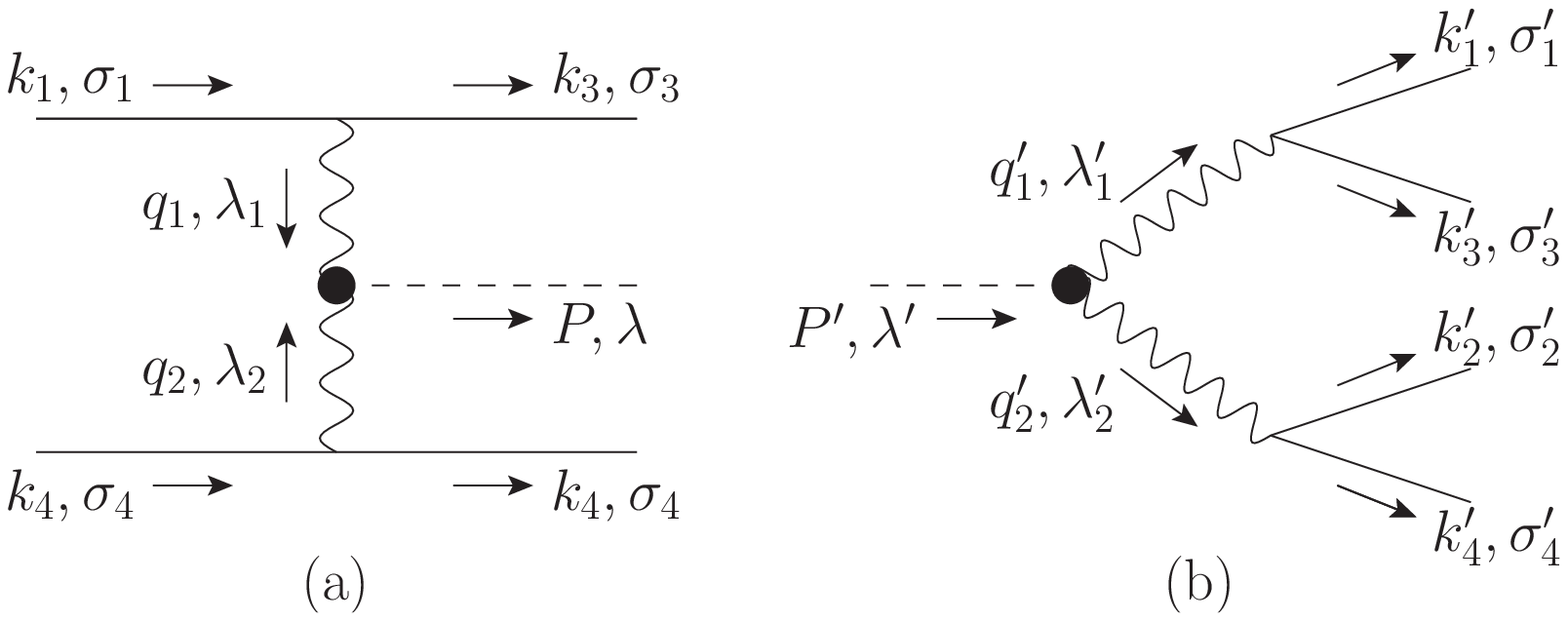,width=.8\textwidth,clip}
 \caption{Schematic view of the subprocesses for (a) the $X$ production
 with 2 jets via VBF, and (b) the $X$ decay to 4 jets via a
 vector-boson pair.  
  The four-momentum and the helicity of each particle are shown. The solid
  lines show either fermions or gluons.
 \label{fig:diagram}}}

Using the completeness relation for space-like vector-bosons ($q_i^2<0$)
\begin{align}
 -g_{\mu'\mu}+\frac{{q_i}_{\mu'}{q_i}_{\mu}}{q_i^2}=
  \sum_{\lam_i=\pm,\,0}(-1)^{\lam_i+1}\,
  \eps_{\mu'}(q_i,\lam_i)^*\,\eps_{\mu}(q_i,\lam_i), 
\label{t-complete}
\end{align}
and neglecting the terms which vanish due to current conservation
\begin{align}
 {q_i}_{\mu}J^{\mu}_{V_ia_ia_{i+2}}(k_i,k_{i+2};\sig_i,\sig_{i+2})=0,
\label{c-conserve}
\end{align}
the VBF helicity amplitudes (\ref{amp}) can be rewritten as the product
of the two incoming current ($f\to fV^*$ or $g\to gV^*$) amplitudes and
the off-shell VBF $X$ production ($V^*V^*\to X$) amplitudes summed over
the polarization of the intermediate vector-bosons 
\begin{align}
 \M^{\lam}_{\sig_1\sig_3,\sig_2\sig_4} = \sum_{V_{1,2}}
  D_{V_1}(q_1^2)D_{V_2}(q_2^2)\sum_{\lam_{1,2}}
  \big(\J^{V_1}_{a_1a_3}\big)^{\lam_1}_{\sig_1\sig_3}
  \big(\J^{V_2}_{a_2a_4}\big)^{\lam_2}_{\sig_2\sig_4}  
  \big(\M^{X}_{V_1V_2}\big)^{\lam}_{\lam_1\lam_2},
\label{pamp}
\end{align}
where
\begin{align}
  \big(\J^{V_i}_{a_ia_{i+2}}\big)^{\lam_i}_{\sig_i\sig_{i+2}}&
 =(-1)^{\lam_i+1}
  J^{\mu}_{V_ia_ia_{i+2}}(k_i,k_{i+2};\sig_i,\sig_{i+2})\,
  \eps_{\mu}(q_i,\lam_i)^*, 
\label{camp}\\
 \big(\M^{X}_{V_1V_2}\big)^{\lam}_{\lam_1\lam_2}&=
  \eps_{\mu_1}(q_1,\lam_1)\,\eps_{\mu_2}(q_2,\lam_2)\,
  \Gamma^{\mu_1\mu_2}_{XV_1V_2}(q_1,q_2;\lam)^*.
\label{VBFamp}
\end{align}
Aside from the summation of $V_{1,2}$, the VBF amplitudes (\ref{pamp})
are generally the coherent sum of the nine amplitudes which have the
different helicity combinations of the colliding vector-bosons.
The explicit forms of the amplitudes (\ref{camp}) and (\ref{VBFamp})
will be given in section~\ref{sec:camp} and \ref{sec:vbfamp}, respectively.
It is worth noting here that the current 
conservation~(\ref{c-conserve}), which ensures that the propagating
vector-bosons have only the three vector-boson components, plays an
essential role in deriving the above expressions, and that it is valid
not only for currents made of massless fermions but also for those made
of on-shell gluons.\\   

In fig.~\ref{fig:diagram}, we notice that the diagrams for (a) the $X$
production with 2 jets via VBF and those for (b) the $X$ decay into 4
jets via a vector-boson pair have identical
topology, even though the intermediate vector-bosons are space-like for
the production while they are time-like for the decay. 
They are related with each other by the crossing symmetry. Therefore, it
may be useful to study the decay angular distributions and correlations
in all the channels simultaneously.

We consider $X$ decays into a (virtual) vector-boson pair which
subsequently decay into $\ell\bar\ell$, $q\bar q$, or $gg$, similar to
the $Xjj$ productions via the VBF processes~(\ref{ff_ffx}),  
\begin{subequations}
\begin{alignat}{3}
 X &\to V^{(*)}V^{(*)} &&\to 
 (\ell\bar\ell/q\bar q)(\ell\bar\ell/q\bar q) 
    &\qquad& (V=W,\,Z,\,\gamma,\,g),\\
 X &\to V^*V^* &&\to (q\bar q)(gg) && (V=g),\\
 X &\to V^*V^* &&\to (gg)(gg) && (V=g),
\end{alignat}
\label{x_jjjj}%
\end{subequations}
and define a common set of kinematical variables as
\begin{align}
 X(P',\lam') &\to {V'_1}^*(q'_1,\lam'_1)+{V'_2}^*(q'_2,\lam'_2) \nn\\
             &\to a'_1(k'_1,\sig'_1)+a'_3(k'_3,\sig'_3)
                 +a'_2(k'_2,\sig'_2)+a'_4(k'_4,\sig'_4),
\label{x_qqqq}
\end{align}
where the same notations for their momenta and helicities are used as in 
the production processes (\ref{qq_qqx}) except for primes ($'$); see also
fig.~\ref{fig:diagram}. 

The decay helicity amplitudes can be expressed
in the same way as the production amplitudes, while 
the completeness relation for time-like vector bosons has to be  
\begin{align}
  -g_{\mu'\mu}+\frac{{q'_i}_{\mu'}{q'_i}_{\mu}}{{q'_i}^2}
 =\sum_{\lam'_i=\pm,\,0}\,
  \eps_{\mu'}(q'_i,\lam'_i)^*\,\eps_{\mu}(q'_i,\lam'_i).
\label{s-complete}
\end{align}
We note that no extra sign factor for $\lam'_i=0$ is needed for
${q'_i}^2>0$. 
The outgoing fermion or gluon current ($V^*\to f\bar f$ or 
$gg$) amplitudes and the $X\to VV$ decay amplitudes,
corresponding to eqs.~(\ref{camp}) and (\ref{VBFamp}) for the
production, are given by 
\begin{align}
   \big(\J'^{V'_i}_{\,a'_ia'_{i+2}}\big)^{\lam'_i}_{\sig'_i\sig'_{i+2}}  
 &=\eps_{\mu}(q'_i,\lam'_i)\,
   J'^{\mu}_{\,V'_ia'_ia'_{i+2}}(k'_i,k'_{i+2};\sig'_i,\sig'_{i+2}), 
\label{dcamp}\\
  \big(\M'^{X}_{\,V'_1V'_2}\big)^{\lam'}_{\lam'_1\lam'_2} &=
  \Gamma^{\mu_1\mu_2}_{XV'_1V'_2}(q'_1,q'_2;\lam')\,
  \eps_{\mu_1}(q'_1,\lam'_1)^*\,\eps_{\mu_2}(q'_2,\lam'_2)^*.
\label{dVBFamp}
\end{align}
The helicity amplitudes for the $X$ decay 
processes~(\ref{x_jjjj}) can now be expressed as
\begin{align}
 \M'^{\lam'}_{\,\sig'_1\sig'_3,\sig'_2\sig'_4} = \sum_{V'_{1,2}}
 D_{V'_1}({q'_1}^2)D_{V'_2}({q'_2}^2)
 \sum_{\lam'_{1,2}}
 \big(\M'^{X}_{\,V'_1V'_2}\big)^{\lam'}_{\lam'_1\lam'_2}
 \big(\J'^{V'_1}_{\,a'_1a'_{3}}\big)^{\lam'_1}_{\sig'_1\sig'_3}
 \big(\J'^{V'_2}_{\,a'_2a'_{4}}\big)^{\lam'_2}_{\sig'_2\sig'_4}.
\label{damp}
\end{align}
Similar to the VBF production amplitudes~(\ref{pamp}), the decay
amplitudes~(\ref{damp}) are generally the sum of the nine amplitudes.
We note our convention that the sign of the current amplitudes with the 
longitudinal polarization is different between the incoming and outgoing 
current amplitudes, (\ref{camp}) and (\ref{dcamp}), reflecting the
difference between (\ref{t-complete}) and (\ref{s-complete}) for the
spin-1 completeness relation.\\

For the sake of our later discussions, we give the complete amplitudes
and the squared matrix elements for the subprocesses of the $X$ plus
$n$-jet production and its subsequent decay into $n'$ jets.
In the vicinity of the $X$ resonance pole, the full amplitudes can be
factorized into the $X$ production amplitudes and its decay
amplitudes, summed over the helicity of the $X$ resonance.
In the $X$ rest frame, if we use the $X$ polarization along the $z$-axis
($J_z=\lambda$) to express the production amplitudes and that of the
decaying direction ($J_{z'}=\lambda'$) to express the decay amplitudes, 
the full amplitudes can be expressed as 
\begin{align}
 \M_{\sig_{1,\cdots,n+2}^{};\,\sig'_{1,\cdots,n'}}
 =\sum_{\lam,\lam'} \M^{\lam}_{\sig_{1,\cdots,n+2}^{}}
  \,D_X(P^2)\,d^{J}_{\lam,\lam'}(\Theta)\,
  \M'^{\lam'}_{\,\sig'_{1,\cdots,n'}}, 
\label{pdamp}
\end{align} 
where $D_{X}(P^2)=(P^2-M^2+iM\Gamma)^{-1}$ times the Wigner's $d$
function $d^{J}_{\lam,\lam'}(\Theta)$ gives the $X$ propagator in the
$X$ rest frame when the $X$ spin is $J$ and the initial and the final
quantization axes have the opening angle $\Theta$. 
For instance, for the $n=0$ case the initial $X$ polarization is 
$\lam=\sig_1-\sig_2$, while the final polarization is 
$\lam'=\sig'_1-\sig'_2$ for $n'=2$.  
We derive the above expression explicitly for the $J=2$ case in
appendix~\ref{sec:dfunc}.
The amplitudes for the scalar particle production and its decay processes
have no $\Theta$ dependence, {\it i.e.}
$d^{J=0}_{\lam,\lam'}(\Theta)=d^{0}_{00}(\Theta)=1$.

It is straightforward to obtain the polarization-summed squared matrix
elements of the full production plus decay amplitudes~(\ref{pdamp}),
\begin{align}
 \sum|\M|^2 &\equiv
 \sum_{\sig_{1,\cdots,n+2}^{}}\sum_{\sig'_{1,\cdots,n'}}
  \big|\M_{\sig_{1,\cdots,n+2}^{};\,\sig'_{1,\cdots,n'}}\big|^2 \nn\\
 &=\big|D_X(P^2)\big|^2
  \sum_{\lam,\lam'}\sum_{\bar\lam,\bar\lam'} 
  \P^{X}_{\lam\bar\lam}\,
   d^{J}_{\lam,\lam'}(\Theta)\,d^{J}_{\bar\lam,\bar\lam'}(\Theta)\,
   \D^{X}_{\lam'\bar\lam'}
\label{M2}
\end{align}
in terms of the production density matrix $\P^X_{\lam\bar\lam}$ and the
decay density matrix $\D^X_{\lam'\bar\lam'}$;
\begin{subequations}
\begin{align}
 \P^{X}_{\lam\bar\lam}&=\sum_{\sigma_{1,\cdots,n+2}^{}}
  \M^{\lam}_{\sig_{1,\cdots,n+2}^{}}
 \big(
  \M^{\bar\lam}_{\sig_{1,\cdots,n+2}^{}}
 \big)^*,\\
 \D^{X}_{\lam'\bar\lam'}&=\sum_{\sigma'_{1,\cdots,n'}}
  \M'{}^{\lam'}_{\sig'_{1,\cdots,n'}}
 \big(
  \M'{}^{\bar\lam'}_{\sig'_{1,\cdots,n'}}
 \big)^*.
\end{align}
\end{subequations}
Although eq.~(\ref{M2}) applies only for parton-level subprocesses, one
can easily generalize it to mixed case and apply it for any processes, 
including summation over different subprocesses and a product of the
relevant parton distribution functions. We note that
the density matrices together with the $d$ functions have all
information on the angular distributions of the final states.

For the VBF processes (\ref{ff_ffx}) and for the $X$ decays into a
pair of vector bosons (\ref{x_jjjj}), the density matrices can be
expressed as
\begin{subequations}
\begin{align}
 \P^X_{\lam\bar\lam}
 &=\sum_{V_{1,2}^{}}
   \big|D_{V_1^{}}(q_1^2)D_{V_2^{}}(q_2^2)\big|^2
   \sum_{\lam_{1,2}}\sum_{\bar\lam_{1,2}}
   \delta(\lam_1-\lam_2-\lam)\,\delta(\bar\lam_1-\bar\lam_2-\bar\lam)\,
   \P^{\lam_1\lam_2}_{\bar\lam_1\bar\lam_2}, 
\label{PtoPtensor}\\
 \D^X_{\lam'\bar\lam'}
 &=\sum_{V'_{1,2}}
   \big|D_{V'_1}({q'_1}^2)D_{V'_2}({q'_2}^2)\big|^2
   \sum_{\lam'_{1,2}}\sum_{\bar\lam'_{1,2}}
   \delta(\lam'_1-\lam'_2-\lam')\,\delta(\bar\lam'_1-\bar\lam'_2-\bar\lam')\,
   \D^{\lam'_1\lam'_2}_{\bar\lam'_1\bar\lam'_2},
\label{DtoDtensor}
\end{align}
\end{subequations}
in terms of the colliding or the decaying vector-boson polarization
amplitudes, (\ref{pamp}) or (\ref{damp}); 
\begin{subequations}
\begin{align}
 \P^{\lam_1\lam_2}_{\bar\lam_1\bar\lam_2}
 &=\sum_{\sigma_{1,\cdots,4}}
 \big[
  \J_1{}^{\lam_1}_{\sig_1\sig_3}
  \J_2{}^{\lam_2}_{\sig_2\sig_4}  
  \M_X{}^{\lam}_{\lam_1\lam_2}
 \big]
 \big[\J_1{}^{\bar\lam_1}_{\sig_1\sig_3}
  \J_2{}^{\bar\lam_2}_{\sig_2\sig_4}  
  \M_X{}^{\bar\lam}_{\bar\lam_1\bar\lam_2}
 \big]^*,
\label{ptensor}\\
 \D^{\lam'_1\lam'_2}_{\bar\lam'_1\bar\lam'_2}
 &=\sum_{\sig'_{1,\cdots,4}}
 \big[
  \M'_X{}^{\lam'}_{\lam'_1\lam'_2}
  \J'_1{}^{\lam'_1}_{\sig'_1\sig'_3}
  \J'_2{}^{\lam'_2}_{\sig'_2\sig'_4}
 \big]
 \big[
  \M'_X{}^{\bar\lam'}_{\bar\lam'_1\bar\lam'_2}
  \J'_1{}^{\bar\lam'_1}_{\sig'_1\sig'_3}
  \J'_2{}^{\bar\lam'_2}_{\sig'_2\sig'_4}
 \big]^*.
\label{dtensor}
\end{align}
\end{subequations}
%

\section{Kinematics}\label{sec:kinematics}

In this section, we define kinematical variables for the production of
the heavy particle $X$ (\ref{qq_qqx}) and its decay (\ref{x_qqqq}).  
The angular configuration of the particles defined below is summarized
in fig.~\ref{fig:frame}. 
It must be noted again that the same notations except for primes ($'$)
are used both for the production and the decay.\\  

First of all, we note the relations between the momenta by the momentum
conservation. 
\begin{align}
 q_1+q_2=P=P'=q'_1+q'_2
\end{align}
with $q_i=k_i-k_{i+2}$ and $q'_i=k'_i+k'_{i+2}$.
On-shell conditions for the external particles and the partonic 
center-of-mass (CM) energy are
\begin{align}
 k_i^2=k_{i+2}^2={k'_i}^2={k'}_{\!\!i+2}^2=0,\quad P^2=P'^2=M^2,\quad
{\rm and}\quad \hat s=(k_1+k_2)^2, 
\label{masslessparton}
\end{align}
while the intermediate vector-bosons are off-shell;
\begin{align}
 q_i^2<0,\quad {q'_i}^2>0.
\end{align}
We would like to express the $VV\to X$ production amplitudes (\ref{VBFamp})
and the $X\to VV$ decay amplitudes (\ref{dVBFamp}) in the $X$ rest frame where
the colliding or decaying (virtual) vector-boson momenta are chosen as
the polar axis, along which their helicities are defined. Because the
helicities are invariant under the boost along the polar axis, we
evaluate the vector-boson emission amplitudes in the Breit frames, 
${q_1^{}}_z=\sqrt{-q_1^2}$ or ${q_2^{}}_z=-\sqrt{-q_2^2}$, and the
vector-boson decay amplitudes in their rest frames, 
${q'_1}^{0}=\sqrt{{q'_1}^2}$ or ${q'_2}^{0}=\sqrt{{q'_2}^2}$.\\

\FIGURE[t]{
 \epsfig{file=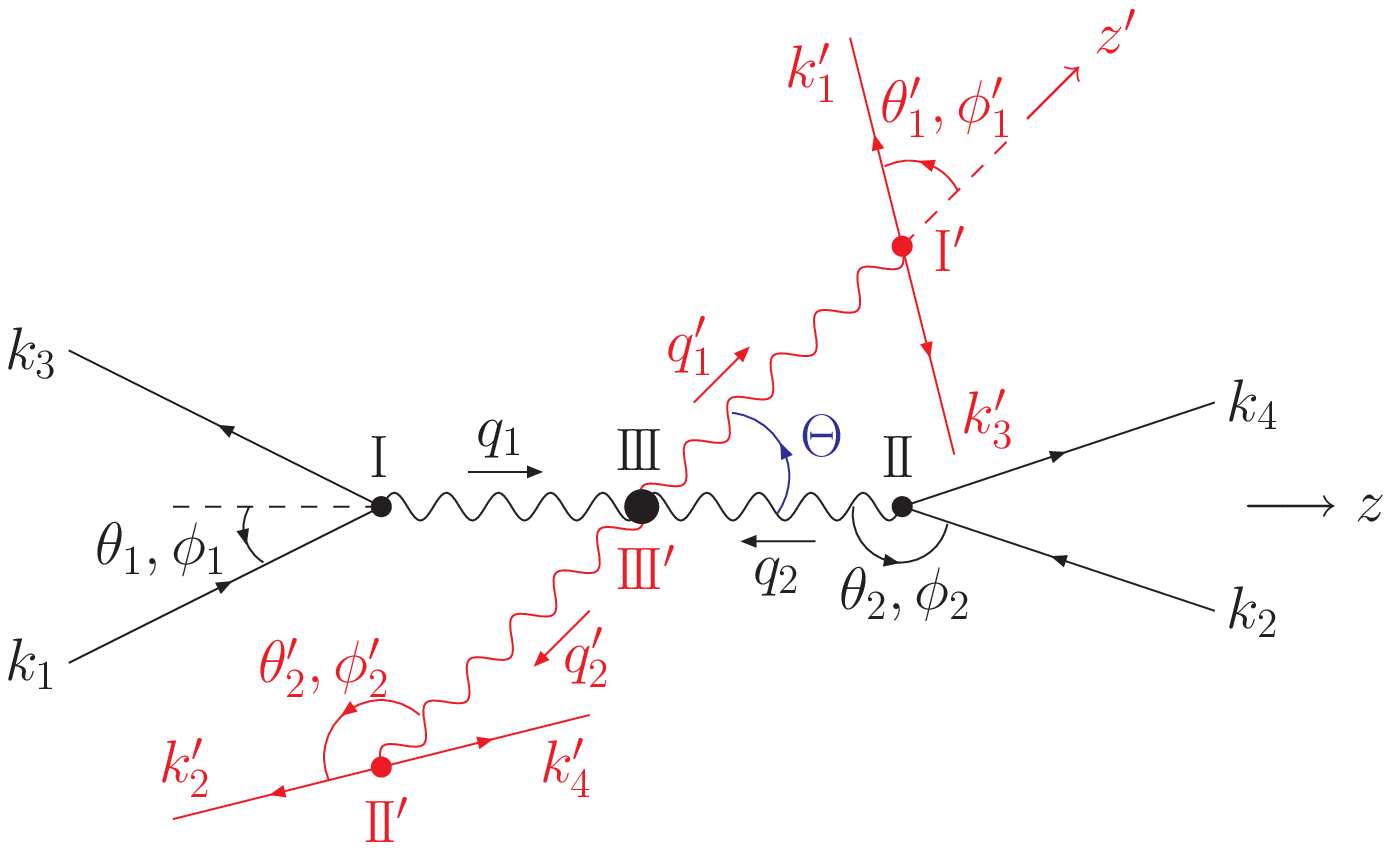,width=.75\textwidth,clip}
 \caption{The momentum and angular configuration of the particles in the
  $q_1$ and $q_2$ Breit frame, (I) and (II), and the VBF frame (III) for
  the production; in the $q'_1$ and $q'_2$ rest frame, (I$'$) and
  (II$'$), and the $X$ rest frame (III$'$) for the decay.
 \label{fig:frame}}} 

For the $X$ + 2-jet production process via VBF (\ref{qq_qqx}), we
parametrize the momenta of the quarks, antiquarks or gluons of the
incoming currents in the Breit frame as follows:
\begin{itemize}
 \item[(I)] the $q_1$ Breit frame
\begin{align}
 q_1^{\mu}&=k_1^{\mu}-k_3^{\mu}=(0,\,0,\,0,\,Q_1), \nn\\ 
 k_1^{\mu}&=\tfrac{Q_1}{2\cos\theta_1}
  (1,\,\sin\theta_1\cos\phi_1,\,\sin\theta_1\sin\phi_1,\,
   \cos\theta_1), \nn\\
 k_3^{\mu}&=\tfrac{Q_1}{2\cos\theta_1}
  (1,\,\sin\theta_1\cos\phi_1,\,\sin\theta_1\sin\phi_1,\,
   -\cos\theta_1),
\label{breit1}
\end{align}
\end{itemize}
where $Q_1=\sqrt{-q_1^2}$, $0<\theta_1<\pi/2$ and $0<\phi_1<2\pi$.  
\begin{itemize}
 \item[(II)] the $q_2$ Breit frame
\begin{align}
 q_2^{\mu}&=k_2^{\mu}-k_4^{\mu}=(0,\,0,\,0,\,-Q_2), \nn\\
 k_2^{\mu}&=-\tfrac{Q_2}{2\cos\theta_2}
  (1,\,\sin\theta_2\cos\phi_2,\,\sin\theta_2\sin\phi_2,\,
   \cos\theta_2), \nn\\
 k_4^{\mu}&=-\tfrac{Q_2}{2\cos\theta_2}
  (1,\,\sin\theta_2\cos\phi_2,\,\sin\theta_2\sin\phi_2,\,
   -\cos\theta_2),
\label{breit2}
\end{align}
\end{itemize}
where $Q_2=\sqrt{-q_2^2}$, $\pi/2<\theta_2<\pi$ and $0<\phi_2<2\pi$.

The momenta of the $t$-channel intermediate vector-bosons are chosen
along the $z$-axis, and the coordinate system where the vector boson has  
a positive (negative) momentum along the $z$-axis is labeled as 1 (2). 
In practice, we always denote by $k_1$ the parton momentum in the
colliding proton with $p_z=\sqrt{s}/2$, while $k_2$ denotes the parton
momentum in the other proton (or anti-proton) with $p_z=-\sqrt{s}/2$, in
the laboratory frame. This agrees with the above definition whenever the
VBF approximation to the full amplitude is valid. The polar angles
$\theta_1$ and $\theta_2$ are measured from the common positive
$z$-axis; see also fig.~\ref{fig:frame}. By a boost along the $z$-axis, 
each Breit frame can be transformed to the rest frame of the heavy
particle, referred to as the VBF frame:  
\begin{itemize}
 \item[(III)] the VBF frame
\begin{align}
 q_1^{\mu}+q_2^{\mu}&=P^{\mu}=P'^{\mu}={q'_1}^{\mu}+{q'_2}^{\mu}
                     =(M,\,0,\,0,\,0), \nn\\
 q_1^{\mu}&=\tfrac{M}{2}
  \big(1-\tfrac{Q_1^2-Q_2^2}{M^2},\,0,\,0,\,\beta\big), \nn\\
 q_2^{\mu}&=\tfrac{M}{2}
  \big(1-\tfrac{Q_2^2-Q_1^2}{M^2},\,0,\,0,\,-\beta\big), \nn\\
 {q'_1}^{\mu} &= \tfrac{M}{2}
  \big(1+\tfrac{{Q'_1}^2-{Q'_2}^2}{M^2},\,\beta'\sin\Theta,\,0,\,
       \beta'\cos\Theta\big), \nn\\
 {q'_2}^{\mu} &= \tfrac{M}{2}
  \big(1+\tfrac{{Q'_2}^2-{Q'_1}^2}{M^2},\,-\beta'\sin\Theta,\,0,\,
       -\beta'\cos\Theta\big),
\label{VBFframe}
\end{align}
\end{itemize}
where $Q'_i=\sqrt{{q'_i}^2}$,
$\beta=\bar\beta\big(-\frac{Q_1^2}{M^2},-\frac{Q_2^2}{M^2}\big)$
and  
$\beta'=\bar\beta\big(\frac{{Q'_1}^2}{M^2},\frac{{Q'_2}^2}{M^2}\big)$
with  
$\bar\beta(a,b)\equiv(1+a^2+b^2-2a-2b-2ab)^{1/2}$, and $\Theta$ is the
angle between the production axis ($z$-axis) and the decay axis
($z'$-axis). The boost factor along the $z$-axis from each Breit frame
to the VBF frame is, respectively, 
\begin{subequations}
\begin{align} 
 \beta_1 &= \big(1-\tfrac{Q_1^2-Q_2^2}{M^2}\big)/\beta, 
\label{boostp1}\\ 
 \beta_2 &= -\big(1-\tfrac{Q_2^2-Q_1^2}{M^2}\big)/\beta. 
\end{align}
\label{boostp}
\end{subequations}

We note that, when the produced particle $X$ decays into
visible particles and its momentum is reconstructed together with those
of the two tagging jets, the Breit frames can in principle be
reconstructed in experiments.\\   

For the $X$ decays into 4 jets via a (virtual) vector-boson pair
(\ref{x_qqqq}), the momenta of the time-like vector-bosons are measured
along the $z'$-axis in the $X$ rest frame: 
\begin{itemize}
 \item[(III$'$)] the $X$ rest frame
\begin{align}
 q_1^{\mu}+q_2^{\mu}&=P^{\mu}=P'^{\mu}={q'_1}^{\mu}+{q'_2}^{\mu}
                     =(M,\,0,\,0,\,0), \nn\\
 q_1^{\mu}&=\tfrac{M}{2}
  \big(1-\tfrac{Q_1^2-Q_2^2}{M^2},\,-\beta\sin\Theta,\,0,\,
       \beta\cos\Theta\big), \nn\\
 q_2^{\mu}&=\tfrac{M}{2}
  \big(1-\tfrac{Q_2^2-Q_1^2}{M^2},\,\beta\sin\Theta,\,0,\,
       -\beta\cos\Theta\big), \nn\\
 {q'_1}^{\mu} &= \tfrac{M}{2}
  \big(1+\tfrac{{Q'_1}^2-{Q'_2}^2}{M^2},\,0,\,0,\,\beta'\big), \nn\\
 {q'_2}^{\mu} &= \tfrac{M}{2}
  \big(1+\tfrac{{Q'_2}^2-{Q'_1}^2}{M^2},\,0,\,0,\,-\beta'\big),
\label{Xrest}
\end{align}
\end{itemize}
which is obtained from the VBF frame (\ref{VBFframe}) by the rotation
with the angle ($-\Theta$) about the $y$-axis.

For the outgoing fermions and gluons, their momenta are parametrized in
the rest frames of the time-like vector-bosons as follows; 
\begin{itemize}
 \item[(I$'$)] the $q'_1$ rest frame: 
\begin{align}
 {q'_1}^{\mu}&={k'_1}^{\mu}+{k'_3}^{\mu}=(Q'_1,\,0,\,0,\,0), \nn\\
 {k'_1}^{\mu}&=\tfrac{Q'_1}{2}(1,\,\sin\theta'_1\cos\phi'_1,\,
                \sin\theta'_1\sin\phi'_1,\,\cos\theta'_1), \nn\\
 {k'_3}^{\mu}&=\tfrac{Q'_1}{2}(1,\,-\sin\theta'_1\cos\phi'_1,\,
                -\sin\theta'_1\sin\phi'_1,\,-\cos\theta'_1),
\label{q1rest}
\end{align}
 \item[(II$'$)] the $q'_2$ rest frame:
\begin{align}
 {q'_2}^{\mu}&={k'_2}^{\mu}+{k'_4}^{\mu}=(Q'_2,\,0,\,0,\,0), \nn\\
 {k'_2}^{\mu}&=\tfrac{Q'_2}{2}(1,\,\sin\theta'_2\cos\phi'_2,\,
                \sin\theta'_2\sin\phi'_2,\,\cos\theta'_2), \nn\\
 {k'_4}^{\mu}&=\tfrac{Q'_2}{2}(1,\,-\sin\theta'_2\cos\phi'_2,\,
                -\sin\theta'_2\sin\phi'_2,\,-\cos\theta'_2),
\label{q2rest}
\end{align}
\end{itemize}
where $0<\theta'_i<\pi$ and $0<\phi'_i<2\pi$. 
In the $X$ rest frame, the $q'_1$ momentum is chosen along the
$z'$-axis, and the $q'_2$ momentum along the negative $z'$-axis.
The polar ($z'$-)axis and the $y$-axis normal to the scattering plane are
chosen common to all the three frames for the decay chain, III$'$,
I$'$ and II$'$, which are related with each other by a boost along the
$z'$-axis; see also fig.~\ref{fig:frame}. For instance, both
$\cos\theta'_1=1$ in (\ref{q1rest}) and $\cos\theta'_2=1$ in
(\ref{q2rest}) denote the momentum along the $z'$-axis direction, and
hence the $a'_1$ momentum ($k'_1$) is along the $V'_1$ momentum while
the $a'_2$ momentum ($k'_2$) is anti-parallel to the $V'_2$ momentum in
the $X$ rest frame. The boost factor along the $z'$-axis from each
vector-boson rest frame to the $X$ rest frame is, respectively, 
\begin{subequations}
\begin{align} 
 \beta'_1 &= \beta'/\big(1+\tfrac{{Q'_1}^2-{Q'_2}^2}{M^2}\big), 
\label{boostd1}\\ 
 \beta'_2 &= -\beta'/\big(1+\tfrac{{Q'_2}^2-{Q'_1}^2}{M^2}\big). 
\end{align}
\label{boostd}
\end{subequations}
%

\section{Helicity amplitudes}\label{sec:helamp}

As we have shown in section~\ref{sec:formalism}, the VBF helicity
amplitudes (\ref{pamp}) can be expressed by the product of the two 
incoming current amplitudes and the $VV\to X$ fusion amplitudes. 
Similarly, the decay helicity amplitudes (\ref{damp}) can be
given by the product of the $X\to VV$ decay amplitudes and the two
outgoing current amplitudes. 
In this section, using the helicity amplitude
technique~\cite{Hagiwara:1985yu} and the kinematical variables
defined in the previous section, we present all the helicity amplitudes
explicitly for the fermion currents, the gluon currents, and the
off-shell VBF vertices, respectively. We also discuss the relation
between the off-shell vector-boson current amplitudes and the standard
parton splitting amplitudes~\cite{Gribov:1972ri}.

\subsection{Current amplitudes}\label{sec:camp}

Let us start with the helicity amplitudes for the incoming fermion
currents ($f\to fV^*$) in the $X$ production process,
$\big(\J^{V_i}_{a_ia_{i+2}}\big)^{\lam_i}_{\sig_i\sig_{i+2}}$ in
eq.~(\ref{camp}). The fermion and antifermion currents are given by 
\begin{align}
 J^{\mu}_{Vff'}(k_i,k_{i+2};\sig_i,\sig_{i+2})
  &= g_{\sig_i}^{Vff'}\,
    \bar u_{f'}(k_{i+2},\sig_{i+2})\,\gamma^{\mu}\,u_f(k_i,\sig_i), \\
 J^{\mu}_{V\bar f\bar f'}(k_i,k_{i+2};\sig_i,\sig_{i+2})
  &= g_{-\sig_i}^{Vff'}\,
    \bar v_f(k_i,\sig_i)\,\gamma^{\mu}\,v_{f'}(k_{i+2},\sig_{i+2}).
\end{align}
Non-vanishing couplings in the standard model (SM) are 
\begin{align}
 &g_{\pm}^{\gamma ff}=eQ_f,\quad 
  g_{\pm}^{gff}=g_st^a, \nn\\
 &g_+^{Zff}=-g_ZQ_f\sin^2\theta_W,\quad 
  g_-^{Zff}=g_Z[T_{3f}-Q_f\sin^2\theta_W], \nn\\
 &g_-^{Wu_id_j}=\big(g_-^{Wd_ju_i}\big)^*=\big(g_W/\sqrt{2}\big)V_{ij},
\label{g_Vff}
\end{align}
where $e=\sqrt{4\pi\alpha}$ is the magnitude of the electron charge,
$g_s=\sqrt{4\pi\alpha_s}$ is the QCD coupling constant, 
$e=g_W\sin\theta_W=g_Z\sin\theta_W\cos\theta_W$, $t^a$ is the
$SU(3)$ color matrix, and $V_{ij}$ denotes the Cabibbo-Kobayashi-Maskawa
(CKM) matrix element. 
Using the kinematical variables defined in the previous section and
contracting the above current with the final-state polarization vector
$\eps^{\mu}(q_i,\lam_i)^*$, we can obtain the helicity amplitudes
explicitly for all the helicity combinations; see eq.~(\ref{camp}). For
our analytical calculations, we use the {\tt HELAS}
convention~\cite{Murayama:1992gi} for the spinors. 
For the virtual vector bosons with space-like momentum,
$q_i^2=-Q_i^2<0$, we define the longitudinal polarization vectors as 
\begin{align}
  \epsilon^{\mu}(q_i^{},\lambda_i^{}=0)
 &=\frac{q_i^0}{Q_i^{}|\vec q_i^{}|}\,(|\vec q_i^{}|^2/q_i^0,\,\vec q_i^{}\,)
\label{pol_l}
\end{align} 
with $Q_i=\sqrt{|(q_i^0)^2-|\vec q_i|^2|}$.
By choosing the transverse polarization vectors 
$\epsilon^{\mu}(q_i,\lambda_i=\pm)$ about the $\vec q_i$ axis, the
identity~(\ref{t-complete}) holds. 

\TABULAR[t]{|l|cc|}{
 \hline\hline \\[-4mm]
  $\Jh_1{}^{\lam_1}_{\sig_1\sig_3}
   (f_{\sig_1}^{}\to f_{\sig_3}^{}V^*_{\lam_1})$ & &
   {\scriptsize $\big[\cos\theta_1\to z_1/(2-z_1)\big]$} \\[0.8mm]   
 \hline \\[-4mm]  
  $\Jh_1{}^+_{++}=-\big(\Jh_1{}^{-}_{--}\big)^*$ & 
   $\dfrac{1}{2\cos\theta_1}(1+\cos\theta_1)\,e^{- i\phi_1}$ & 
   $\dfrac{1}{z_1}\,e^{- i\phi_1}$ \\[2mm]
  $\Jh_1{}^0_{++}=\Jh_1{}^0_{--}$ & 
   $-\dfrac{1}{\sqrt{2}\cos\theta_1}\sin\theta_1$ & 
   $-\dfrac{\sqrt{2(1-z_1)}}{z_1}$ \\[2mm] 
  $\Jh_1{}^{-}_{++}=-\big(\Jh_1{}^{+}_{--}\big)^*$ & 
   $-\dfrac{1}{2\cos\theta_1}(1-\cos\theta_1)\,e^{ i\phi_1}$ & 
   $-\dfrac{1-z_1}{z_1}\,e^{ i\phi_1}$ \\[3mm]
  $\Jh_1{}^{\lam_1}_{+-}=\Jh_1{}^{\lam_1}_{-+}$ & $0$ & $0$ 
 \\[0.8mm] 
 \hline\hline \\[-4mm]
  $\Jh'_1{}^{\lam'_1}_{\sig'_1\sig'_3}
   (V^*_{\lam'_1}\to f_{\sig'_1}^{}\bar f_{\sig'_3}^{})$ & &  
   {\scriptsize $\big[\cos\theta'_1\to 2z'_1-1\big]$} \\[1.6mm]
 \hline \\[-4mm]
  $\Jh'_1{}^{+}_{+-}=-\big(\Jh'_1{}^{-}_{-+}\big)^*$ &
   $\dfrac{1}{2}(1+\cos\theta'_1)\,e^{ i\phi'_1}$ & 
   $z'_1\,e^{ i\phi'_1}$ \\
  $\Jh'_1{}^0_{+-}=\Jh'_1{}^0_{-+}$ & 
   $\dfrac{1}{\sqrt{2}}\sin\theta'_1$ & 
   $\sqrt{2z'_1(1-z'_1)}$ \\
  $\Jh'_1{}^{-}_{+-}=-\big(\Jh'_1{}^{+}_{-+}\big)^*$ &
   $\dfrac{1}{2}(1-\cos\theta'_1)\,e^{- i\phi'_1}$ & 
   $(1-z'_1)\,e^{- i\phi'_1}$ \\[2.5mm]
  $\Jh'_1{}^{\lam'_1}_{++}=\Jh'_1{}^{\lam'_1}_{--}$ & $0$ & $0$
 \\[1mm] 
 \hline}
{\label{Jq} The reduced helicity amplitudes for the off-shell
 vector-boson currents: the incoming fermion currents  
 $\Jh_1{}^{\lam_1}_{\sig_1\sig_3}(f\to fV^*)$ in the Breit frame
 (top), and  the outgoing fermion currents 
 $\Jh'_1{}^{\lam'_1}_{\sig'_1\sig'_3}(V^*\to f\bar f)$ in the
 vector-boson rest frame (bottom).
 In the third column the splitting amplitudes are also shown in the
 collinear limit, where $z_1^{(\prime)}$ is the energy fraction of the
 initial particle.}

Here, for notational convenience, we define the reduced current
amplitudes $\Jh_i{}^{\lam_i}_{\sig_i\sig_{i+2}}$ as   
\begin{align}
 \big(\J^{V_i}_{a_ia_{i+2}}\big)^{\lam_i}_{\sig_i\sig_{i+2}}
  =\sqrt{2}\,g_{\sig_i}^{V_ia_ia_{i+2}}Q_i\,
   \Jh_i{}^{\lam_i}_{\sig_i\sig_{i+2}}.
\label{rtamp}
\end{align}
In table~\ref{Jq}(top), the reduced helicity amplitudes for the incoming
fermion currents, $\Jh_1{}^{\lam_1}_{\sig_1\sig_3}$ 
$(f_{\sig_1}^{}\to f_{\sig_3}^{}V^*_{\lam_1})$, are shown in the $q_1$
Breit frame (\ref{breit1}), or the frame I in fig.~\ref{fig:frame}. 
The following features of the amplitudes are worth noting:  
(i) The reduced amplitudes for the antiquark currents are the same as
those for the quark currents. 
(ii) Parity transformation gives the relation 
$\Jh_1{}^{\lam_1}_{\sig_1,\sig_3}=
 (-1)^{\lam_1}(\Jh_1{}^{-\lam_1}_{-\sig_1,-\sig_3})^*$.  
(iii) The quark masses are neglected (eq.~(\ref{masslessparton})), 
and hence the helicity-flip
amplitudes $\Jh_1{}^{\lam_1}_{\sig,-\sig}$ are zero due to the
chirality conservation.   
(iv) The amplitudes $\Jh_2$ are related to $\Jh_1$ by
\begin{align} 
  \Jh_2{}^{\lam_2}_{\sig_2\sig_4}(\theta_2,\phi_2)
 =(-1)^{\lambda_2+1}\Jh_1{}^{-\lam_2}_{\sig_2\sig_4}(\theta_2,\phi_2).
\label{J2in_con}
\end{align}
(v) The $1/\cos\theta_1$ dependence comes from the common factor of the
four-momentum in the $q_1$ Breit frame (\ref{breit1}), and this gives
rise to the enhancement of the amplitudes when $\cos\theta_1$ approaches
zero. This singularity simply reflects the well-known soft singularity
in the laboratory frame, and will be discussed further in 
section~\ref{sec:split}. 
(vi) The transverse currents $\Jh^{\pm}$ have opposite phases with
each other in terms of the azimuthal angle, while the longitudinal
currents $\Jh^{0}$ do not have the azimuthal angle dependence.

Next, for comparison, we consider the helicity amplitudes for the
outgoing fermion currents ($V^*\to f\bar f$) in the $X$ decay process, 
$\big(\J'^{\,V'_i}_{\ a'_ia'_{i+2}}\big)^{\lam'_i}_{\sig'_i\sig'_{i+2}}$
in eq.~(\ref{dcamp}). 
The fermion currents in which the time-like vector-bosons decay into
$\ell\bar\ell$ or $q\bar q$ are given by
\begin{align}
  J'^{\,\mu}_{\ Vf\bar f'}(k'_i,k'_{i+2};\sig'_i,\sig'_{i+2})
 =g_{\sig'_i}^{Vff'}\,
  \bar u_f(k'_i,\sig'_i)\,\gamma^{\mu}\,v_{f'}(k'_{i+2},\sig'_{i+2}).
\label{dcurrent}
\end{align}
The reduced amplitudes for the outgoing currents, defined as in
eq.~(\ref{rtamp}), 
\begin{align}
  \big(\J'^{\,V'_i}_{\ a'_ia'_{i+2}}\big)^{\lam'_i}_{\sig'_i\sig'_{i+2}}
 =\sqrt{2}\,g_{\sig'_i}^{V'_ia'_ia'_{i+2}}Q'_i\,
   \Jh'_i{}^{\lam'_i}_{\sig'_i\sig'_{i+2}},
\end{align}
can be obtained by contracting the current (\ref{dcurrent}) with the
initial-state polarization vector $\eps^{\mu}(q'_i,\lam'_i)$; see
eq.~(\ref{dcamp}). 

In table~\ref{Jq}(bottom), the reduced helicity amplitudes for the
outgoing fermion currents, 
$\Jh'_1{}^{\lam'_1}_{\sig'_1\sig'_3}
 (V^*_{\lam'_1}\to f_{\sig'_1}^{}\bar f_{\sig'_3}^{})$, 
are shown in the $q'_1$ rest frame (\ref{q1rest}), or the frame I$'$ in
fig.~\ref{fig:frame}.
The amplitudes have similar features to the incoming current amplitudes
in table~\ref{Jq}(top).  
It is worth noting that not only the $s$-channel amplitudes 
$\Jh'_1{}^{\lam'_1}_{\sig'_1\sig'_3}(V^*\to f\bar f)$ but also the
$t$-channel ones $\Jh_1{}^{\lam_1}_{\sig_1\sig_3}(f\to fV^*)$ can be
expressed by the same $J=1$ $d$~functions as
\begin{align} 
 \Jh_1{}^{\lam_1}_{\sig_1\sig_3}(f\to fV^*) &\propto 
  d^{\,1}_{\lam_1,\,\sig_1+\sig_3}(\theta_1), \\
 \Jh'_1{}^{\lam'_1}_{\sig'_1\sig'_3}(V^*\to f\bar f) &\propto 
  d^{\,1}_{\lam'_1,\,\sig'_1-\sig'_3}(\theta'_1).
\end{align}
The different points from the incoming current amplitudes are the
following:  
(i) There is no $1/\cos\theta_1$ factor.
(ii) The phases have opposite
signs, reflecting the wavefunctions of the incoming and outgoing
vector-bosons. 
(iii) The same-helicity amplitudes $\Jh'_1{}^{\lam'_1}_{\sig',\sig'}$
are zero, which correspond to the chirality-flip amplitudes. 
(iv) The amplitudes $\Jh'_2$ are given by
\begin{align}   
   \Jh'_2{}^{ \lam'_2}_{\sig'_2\sig'_4}(\theta'_2,\phi'_2)
 =-\Jh'_1{}^{-\lam'_2}_{\sig'_2\sig'_4}(\theta'_2,\phi'_2).
\label{J2qout_con}
\end{align}

It is worth pointing out here that the list of the current amplitudes in
table~\ref{Jq} is useful not only for hadron colliders but
also for $e^+e^-$, $ep$ and $\gamma\gamma$ colliders.\\

Turning now to the gluon current amplitudes, the incoming gluon currents
in the $X$ production process, $g\to gV^*$, where $V^*$ is a virtual
gluon, are given by 
\begin{multline}
 J^{\mu}_{Vgg}(k_i,k_{i+2};\sig_i,\sig_{i+2})
  =g_sf^{abc}\,\eps_{\alp}^b(k_i,\sig_i)\,
    \eps_{\bet}^{c}(k_{i+2},\sig_{i+2})^* \\
  \quad\times\big[-g^{\alp\bet}(k_i+k_{i+2})^{\mu}
                  -g^{\bet\mu}(-k_{i+2}+q_i)^{\alp}
                  -g^{\mu\alp}(-q_i-k_i)^{\bet}
             \big],
\label{gcurrent}
\end{multline}
where $f^{abc}$ is the structure constant of the $SU(3)$ group.  
Similar to the fermion currents, the reduced amplitudes are defined as
\begin{align}
  \big(\J^{V}_{gg}\big)^{\lam_i}_{\sig_i\sig_{i+2}}
 &=(-1)^{\lam_i+1}
  J^{\mu}_{Vgg}(k_i,k_{i+2};\sig_i,\sig_{i+2})\,
  \eps_{\mu}(q_i,\lam_i)^* \nn\\
 &=\sqrt{2}\,g_sf^{abc}\,Q_i\,
   \Jh_i{}^{\lam_i}_{\sig_i\sig_{i+2}}.
\label{camp_gluon}
\end{align}
For the polarization vectors of the external gluons in the  
amplitude~(\ref{camp_gluon}), we adopt a common light-cone gauge, 
$n_i^{\mu}=(1,-\vec q_i/|\vec q_i|)$, which satisfy
\begin{align}
  n_i\cdot\epsilon(k_{i},\sig_{i})
 =n_i\cdot\epsilon(k_{i+2},\sig_{i+2})
 =n_i^2=0,\quad n_i\cdot k_i\ne 0,
 \quad n_i\cdot k_{i+2}\ne 0.
\end{align}
It should be noted that these gauge-fixing vectors are boost invariant
along the current momentum directions, in particular between the Breit
frames of I and II for $i=1$ and $2$, respectively, and the collision CM
frame III. 

By the crossing symmetry, the outgoing gluon currents in the $X$ decay
process, $V^*\to gg$, are obtained by making the replacements in
eq.~(\ref{gcurrent}): $k_i\to -k'_i$, $k_{i+2}\to k'_{i+2}$, 
$q_i\to -q'_i$, and $\eps_{\alp}\to\eps_{\alp}^*$.   

\TABULAR[t]{|l|cc|}{
 \hline\hline \\[-4mm]
  $\Jh_1{}^{\lam_1}_{\sig_1\sig_3}
   (g_{\sig_1}^{}\to g_{\sig_3}^{}V^*_{\lam_1})$ & & 
   {\scriptsize 
    $\big[\cos\theta_1\to z_1/(2-z_1)\big]$}
 \\[0.8mm] 
 \hline \\[-4mm]
  $\Jh_1{}^{+}_{++}=-\big(\Jh_1{}^{-}_{--}\big)^*$ &
   $\dfrac{1}{2\sin\theta_1\cos\theta_1}(1+\cos\theta_1)^2\,
    e^{- i\phi_1}$ &
   $\dfrac{1}{z_1\sqrt{1-z_1}}\,e^{- i\phi_1}$ \\[2.5mm]
  $\Jh_1{}^0_{++}=\Jh_1{}^0_{--}$ & 
   $-\dfrac{1}{\sqrt{2}\cos\theta_1}$ & 
   $-\dfrac{2-z_1}{\sqrt{2}\,z_1}$ \\[2.5mm] 
  $\Jh_1{}^{-}_{++}=-\big(\Jh_1{}^{+}_{--}\big)^*$ &
   $-\dfrac{1}{2\sin\theta_1\cos\theta_1}(1-\cos\theta_1)^2\,
    e^{ i\phi_1}$ &
   $-\dfrac{(1-z_1)^2}{z_1\sqrt{1-z_1}}\,e^{ i\phi_1}$ \\[2.5mm]
  $\Jh_1{}^{+}_{+-}=-\big(\Jh_1{}^{-}_{-+}\big)^*$ & 
   $-\dfrac{2}{\tan\theta_1}\,e^{i\phi_1}$ & 
   $-\dfrac{z_1}{\sqrt{1-z_1}}\,e^{i\phi_1}$ \\[4mm]
  $\Jh_1{}^{0/-}_{+-}=\Jh_1{}^{0/+}_{-+}$ & $0$ & $0$
 \\[1.3mm] 
 \hline\hline \\[-4mm]
  $\Jh'_1{}^{\lam'_1}_{\sig'_1\sig'_3}
   (V^*_{\lam'_1}\to g_{\sig'_1}^{}g_{\sig'_3}^{})$ & & 
   {\scriptsize 
    $\big[\cos\theta'_1\to 2z'_1-1\big]$} 
 \\[1.6mm]
 \hline \\[-4mm]
  $\Jh'_1{}^{+}_{+-}=-\big(\Jh'_1{}^{-}_{-+}\big)^*$ &
   $-\dfrac{1}{2\sin\theta'_1}(1+\cos\theta'_1)^2\,e^{ i\phi'_1}$ &
   $-\dfrac{{z'_1}^2}{\sqrt{z'_1(1-z'_1)}}\,e^{ i\phi'_1}$ \\
  $\Jh'_1{}^0_{+-}=\Jh'_1{}^0_{-+}$ & 
   $-\dfrac{1}{\sqrt{2}}\cos\theta'_1$ & 
   $-\dfrac{2z'_1-1}{\sqrt{2}}$ \\ 
  $\Jh'_1{}^{-}_{+-}=-\big(\Jh'_1{}^{+}_{-+}\big)^*$ &
   $\dfrac{1}{2\sin\theta'_1}(1-\cos\theta'_1)^2\,e^{- i\phi'_1}$ &
   $\dfrac{(1-z'_1)^2}{\sqrt{z'_1(1-z'_1)}}\,e^{- i\phi'_1}$ \\
  $\Jh'_1{}^{+}_{++}=-\big(\Jh'_1{}^{-}_{--}\big)^*$ &
   $\dfrac{2}{\sin\theta'_1}\,e^{-i\phi'_1}$ & 
   $\dfrac{1}{\sqrt{z'_1(1-z'_1)}}\,e^{-i\phi'_1}$ \\[4mm]
  $\Jh'_1{}^{0/-}_{++}=\Jh'_1{}^{0/+}_{--}$ & $0$ & $0$
 \\[1.3mm]  
 \hline}
{\label{Jg} The same as table \ref{Jq}, but for the gluon currents, 
 $g\to gV^*$ (top) and $V^*\to gg$ (bottom), where $V^*$ is an off-shell
 gluon.} 

In table~\ref{Jg}, we present the reduced helicity amplitudes for the
incoming gluon currents,
$\Jh_1{}^{\lam_1}_{\sig_1\sig_3}
 (g_{\sig_1}^{}\to g_{\sig_3}^{}V^*_{\lam_1})$, 
in the $q_1$ Breit frame (top), and for the outgoing gluon currents, 
$\Jh'_1{}^{\lam'_1}_{\sig'_1\sig'_3}
 (V^*_{\lam'_1}\to g_{\sig'_1}^{}g_{\sig'_3}^{})$, 
in the $q'_1$ rest frame (bottom).
The derivations are straightforward as in the fermion case and the
amplitudes have similar properties to those for fermions in
table~\ref{Jq}. However, the results for the gluon currents are more
involved.  
The incoming amplitudes $\Jh_2$ for the helicity-conserved currents 
($\sigma_2=\sigma_4$) are given by eq.~(\ref{J2in_con}), while the
outgoing ones $\Jh'_2$ for the opposite-helicity currents 
($\sigma'_2=-\sigma'_4$) are obtained by 
\begin{align}
  \Jh'_2{}^{ \lam'_2}_{\sig'_2\sig'_4}(\theta'_2,\phi'_2)
 =\Jh'_1{}^{-\lam'_2}_{\sig'_2\sig'_4}(\theta'_2,\phi'_2).
\label{J2gout_con}
\end{align}
Unlike the fermion current amplitudes, some of the helicity-flip
amplitudes ($\sigma_1=-\sigma_3$) for the incoming currents and the 
same-helicity amplitudes ($\sigma'_1=\sigma'_3$) for the outgoing
currents are nonzero, and the amplitudes $\Jh_2^{(\prime)}$ are
given by 
\begin{align}
  \Jh^{(\prime)}_2{}^{\lam_2^{(\prime)}}_{\sig_2^{(\prime)}\sig_4^{(\prime)}}(\theta^{(\prime)}_2,\phi^{(\prime)}_2)
 =-\Jh^{(\prime)}_1{}^{\lam^{(\prime)}_2}_{\sig^{(\prime)}_2\sig^{(\prime)}_4}(\theta^{(\prime)}_2,\phi^{(\prime)}_2).
\end{align}
Furthermore, in addition to the singularity of the amplitudes at
$\cos\theta_1=0$, which appears also in the incoming fermion amplitudes
in table~\ref{Jq}(top), the singularity at $\sin\theta_1^{(\prime)}=0$,
or $\cos\theta_1^{(\prime)}=1$ also exists; see more discussions in
section~\ref{sec:split}. 
It must be stressed here that the phase dependence of the gluonic
currents and that of the fermionic currents are very similar,
$e^{-i\lambda_1\phi_1}$ or $e^{i\lambda'_1\phi'_1}$ when the gluon
helicity is conserved ($q_i^2<0$) or flipped (${q'}_i^2>0$) as in the
fermionic case. These phases lead to the azimuthal angle
correlations of the jets, as we will show later.

\subsection{Relation to the splitting amplitudes}\label{sec:split}

Before turning to the $XVV$ amplitudes, it may be valuable to discuss
the off-shell vector-boson current amplitudes from a different point of 
view, parton branching description~\cite{Gribov:1972ri}, where the
outgoing particles are emitted collinearly.\\   

To begin with, we consider the incoming current amplitudes ($f\to fV^*$
or $g\to gV^*$).
Let $z$ be the energy fraction of the initial parton that is carried off
by the space-like vector-boson.  
In the VBF frame (\ref{VBFframe}), the energy fraction  
$z_1=q_1^0/k_1^0$ is written in terms of $\cos\theta_1$ defined
in the $q_1$ Breit frame (\ref{breit1}) and the boost factor $\beta_1$
in eq.~(\ref{boostp1}) as   
\begin{align}
 z_1 = \frac{q_1^0}{k_1^0} 
     = \frac{2\beta_1\cos\theta_1}{1+\beta_1\cos\theta_1}. 
\end{align}
Taking the $\beta_1=1$ limit, where the space-like vector-boson becomes
on-shell and collinear with the final parton, we obtain the simple
relation between the Breit-frame angle $\cos\theta_1$ and the energy
fraction $z_1$ as  
\begin{align}
 \cos\theta_1 = \frac{z_1}{\beta_1(2-z_1)}\
  \stackrel{\bet_1=1}{\longrightarrow}\ \frac{z_1}{2-z_1}.
\label{cos1z1t}
\end{align}

In the third column in tables~\ref{Jq}(top) and \ref{Jg}(top), using
the above relation (\ref{cos1z1t}) in the $\beta_1=1$ limit, or in the
collinear limit, the helicity amplitudes for the incoming fermion and
gluon currents are rewritten as splitting amplitudes with appropriate
phases. These formulae may give us clear explanation for the
origin of the singularities of the amplitudes which we encountered in
the previous section. We see from table~\ref{Jq}(top) for the incoming
fermion splitting that the amplitudes are enhanced at $z_1\to 0$, where
the vector boson becomes soft.
On the other hand, for the $g\to gV^*$ splitting in table~\ref{Jg}(top),
the enhancements of the amplitudes at $z_1\to 0$ and 1 are associated
with the soft emissions of the space-like gluon and the outgoing jet
gluon, respectively.\\

Next, we consider the outgoing current amplitudes ($V^*\to f\bar f$ or
$gg$). Here, we define a fraction $z'$ as the energy transferred from
the time-like vector-boson to the outgoing fermion or gluon.
In the $X$ rest frame (\ref{Xrest}),   
\begin{align}
 z'_1 = \frac{{k'_1}^0}{{q'_1}^0} 
     = \frac{1}{2\beta'_1}(1+\beta'_1\cos\theta'_1),
\end{align}
where $\cos\theta'_1$ is defined in the $q'_1$ rest frame (\ref{q1rest})
and $\beta'_1$ is the boost factor in eq.~(\ref{boostd1}). In the
$\beta'_1=1$ limit, where the time-like vector-boson becomes on
mass-shell and the two outgoing partons are emitted collinearly, we obtain 
\begin{align}
 \cos\theta'_1 = 2z'_1-\frac{1}{\beta'_1}\
  \stackrel{\beta'_1=1}{\longrightarrow}\ 2z'_1-1.
\label{cos1z1s}
\end{align}

In the third column in tables~\ref{Jq}(bottom) and \ref{Jg}(bottom),
by making the replacement (\ref{cos1z1s}) in the $\beta'_1=1$
limit, we present the splitting amplitudes for the outgoing fermion and
gluon currents. There is no singularity for the $V^*\to f\bar f$
splitting amplitudes, while the $V^*\to gg$ amplitudes have the
singularities at $z'_1=0$ and 1, similar to the space-like gluon
splitting amplitudes. This is because the singularities are associated
only with soft gluon emissions.\\ 

Finally, let us confirm that the splitting amplitudes discussed above
reproduce the standard (unregularized) quark and gluon splitting
functions. 
From table~\ref{Jq}(top), the sum of the squared amplitudes for the
transversely-polarized vector-bosons ($\lam_1=\pm1$), averaged/summed
over initial/final parton helicities, leads to the $f\to V_T$
splitting function as 
\begin{align}
 z_1\,\frac{1}{2}\sum_{\sig_{1,3}=\pm} 
  \big[ |\Jh_1{}^{+}_{\sig_1\sig_3}|^2
       +|\Jh_1{}^{-}_{\sig_1\sig_3}|^2 \big]
  = \frac{1+(1-z_1)^2}{z_1} = P_{V_T/f}(z_1),
\end{align}
where the extra factor $z_1$ comes from the initial-state flux factor
for the space-like branching. Similarly, from table~\ref{Jg}(top) we can 
obtain the gluon splitting function as
\begin{align}
 z_1\,\frac{1}{4}\sum_{\sig_{1,3}=\pm} 
   \big[ |\Jh_1{}^{+}_{\sig_1\sig_3}|^2
        +|\Jh_1{}^{-}_{\sig_1\sig_3}|^2 \big]
 = \frac{1-z_1}{z_1}+\frac{z_1}{1-z_1}+z_1(1-z_1)
 = P_{V_T/V_T}(z_1),
\label{gsplit}  
\end{align}
where, in addition to the spin averaged factor, we divide by the
statistical factor for the two identical gluons in the final state. 
On the other hand, from the $V^*\to f\bar f$ splitting amplitudes in
table~\ref{Jq}(bottom), which are time-like branching, the $V_T\to f$
splitting function can be reproduced, 
\begin{align}
 \frac{1}{2}\sum_{\sig'_{1,3}=\pm} 
   \big[ |\Jh'_1{}^{+}_{\sig'_1\sig'_3}|^2
        +|\Jh'_1{}^{-}_{\sig'_1\sig'_3}|^2 \big]
 = {z'_1}^2+(1-{z'_1}^2) 
 = P_{f/V_T}(z'_1). 
\label{PVTq}
\end{align}
Likewise, from table~\ref{Jg}(bottom), we can obtain the gluon splitting 
function as in eq.~(\ref{gsplit}), 
\begin{align}
 \frac{1}{4}\sum_{\sig'_{1,3}=\pm} 
   \big[ |\Jh'_1{}^{+}_{\sig'_1\sig'_3}|^2
        +|\Jh'_1{}^{-}_{\sig'_1\sig'_3}|^2 \big]
 = P_{V_T/V_T}(z'_1).
\label{PVTVT}
\end{align}

In the parton branching description, the space-like and time-like
vector-bosons, {\it i.e.} gluons, are almost on mass-shell, and hence
their polarization vectors are taken to be purely transverse.
However, the longitudinal component of the polarization ($\lam_1=0$)
also exists for the massive vector-bosons.
Therefore, in addition to the above standard parton
splitting functions for the transversely-polarized vector-bosons, we list
the functions for the longitudinal polarization:  
\begin{align}
 z_1\,\frac{1}{2}\sum_{\sig_{1,3}=\pm}|\Jh_1{}^{0}_{\sig_1\sig_3}|^2
  &= \frac{2(1-z_1)}{z_1} = P_{V_L/f}(z_1)
  &&\text{from table~\ref{Jq}(top)}, \\
 z_1\,\frac{1}{4}\sum_{\sig_{1,3}=\pm}|\Jh_1{}^{0}_{\sig_1\sig_3}|^2
  &= \frac{(2-z_1)^2}{4z_1} = P_{V_L/V_T}(z_1)
  &&\text{from table~\ref{Jg}(top)}, \\
 \sum_{\sig'_{1,3}=\pm}|\Jh'_1{}^{0}_{\sig'_1\sig'_3}|^2
  &= 4z'_1(1-z'_1) = P_{f/V_L}(z'_1)
  &&\text{from table~\ref{Jq}(bottom)}, \label{PqVL}\\
 \frac{1}{2}\sum_{\sig'_{1,3}=\pm}|\Jh'_1{}^{0}_{\sig'_1\sig'_3}|^2
  &= \frac{1}{2}(2z'_1-1)^2 = P_{V_T/V_L}(z'_1)
  &&\text{from table~\ref{Jg}(bottom)}. \label{PVLVT}
\end{align}
In fact, many studies have been performed for Higgs boson productions via WBF
in the equivalent weak-boson approximation, where the $t$-channel
intermediate weak-bosons are viewed as partons in the incoming quarks and
the above splitting functions $P_{V_L/f}$ as well as 
$P_{V_T/f}$ are considered~\cite{Cahn:1983ip}; see more details
in ref.~\cite{Djouadi:2005gi} and references therein.

\subsection{Off-shell VBF amplitudes}\label{sec:vbfamp}

We will now show the final piece, the $VV\to X$ production and the
$X\to VV$ decay amplitudes in eqs.~(\ref{VBFamp}) and (\ref{dVBFamp}),
respectively. 
In this paper, we consider the productions and the decays of massive spin-0
and spin-2 bosons: neutral $CP$-even and $CP$-odd Higgs bosons ($X=H$
and $A$), and graviton resonances ($X=G$). 

For the fusion vertex of a $CP$-even Higgs boson, we consider both the 
WBF process and the GF process through a top-quark loop in the SM, and
their tensor structures $\Gamma^{\mu_1\mu_2}_{XV_1V_2}(q_1,q_2)$, normalized
by the coupling form factors $g_{XV_1V_2}^{}(q_1,q_2)$, are given in
table~\ref{VVX}. The constant coupling, $g_{HVV}^{}=2m_V^2/v\ (V=W,Z)$ with
$v=246$ GeV, gives the WBF vertex, while the
explicit expression of the form factor $g_{Hgg}^{}(q_1,q_2)$ by a
triangle-loop is given in ref.~\cite{DelDuca:2001eu}.%
\footnote{Strictly speaking, the $Hgg$ coupling tensor is given by two
terms as  
 $T^{\mu_1\mu_2}=F_T^{}T_T^{\mu_1\mu_2}
 +F_L^{}T_L^{\mu_1\mu_2}$~\cite{DelDuca:2001eu}, where $T_T^{\mu_1\mu_2}$
 is identical with the form in table~\ref{VVX}.
 However, we neglect the second term since it does not contribute in the 
 on-shell gluon limit and is not enhanced in the collinear limit.}
Note that in 
refs.~\cite{Plehn:2001nj,Hankele:2006ja,Figy:2004pt,Hankele:2006ma}
the same loop-induced vertex structure has been considered to study the
anomalous couplings between the Higgs and weak bosons.
We also consider the GF vertex for a $CP$-odd Higgs boson, defined in
table~\ref{VVX}.  
For light Higgs bosons ($M_{H,A}<2m_t$), the above $Hgg$ and $Agg$
vertices can be well described by the heavy-top effective 
Lagrangian~\cite{Dawson:1990zj,Spira:1995rr,Kauffman:1996ix,Kauffman:1998yg}
\begin{align}
 {\cal L}_{H,A}=-\frac{1}{4}g_{Hgg}^{}HF_{\mu\nu}^aF^{a,\mu\nu}
                +\frac{1}{2}g_{Agg}^{}AF_{\mu\nu}^a\tilde F^{a,\mu\nu},
\end{align}
where $F^{a,\mu\nu}$ is the gluon field-strength tensor and 
$\tilde F^{a,\mu\nu}
 =\frac{1}{2}\eps^{\mu\nu\rho\sig}F_{\rho\sig}^a$ is its dual. 
The coupling constants are given by 
$g_{Hgg}^{}=\alpha_s g_{Htt}^{}/3\pi m_t$
and $g_{Agg}^{}=\alpha_s g_{Att}^{}/2\pi m_t$. 
The same tensor structures can be written for the interactions of the
$CP$-even/odd Higgs boson with two photons~\cite{Shifman:1979eb} and
with a $Z$-boson and a photon~\cite{Cahn:1978nz}.   

\TABULAR[t]{|ccc|l|}{\hline\\[-5mm]
  $X$ & $(\lam)$ & $V_i$ & 
  $\Gamma^{\mu_1\mu_2}_{XV_1V_2}(q_1,q_2;\lam)/g_{XV_1V_2}^{}(q_1,q_2)$  
 \\[0.5mm] \hline\\[-4mm]
  $H$ & $(0)$ & $W,Z$ & $g^{\mu_1\mu_2}$ \\
  $H$ & $(0)$ & $\gamma,Z/\gamma,g$ &
   $q_1\cdot q_2\,g^{\mu_1\mu_2}-q_2^{\mu_1}q_1^{\mu_2}$ \\
  $A$ & $(0)$ & $\gamma,Z/\gamma,g$ & 
   $\eps^{\mu_1\mu_2\alpha\beta}{q_1}_{\alpha}{q_2}_{\beta}$ \\
  $G$ & $(\pm2,\pm1,0)$ & $W,Z,\gamma,g$ & 
   $\eps_{\alpha\beta}(P,\lam)\,
    \hat\Gamma^{\alpha\beta,\mu_1\mu_2}_{GVV}(q_1,q_2)$ \\[1mm] 
 \hline}
{\label{VVX} The $XVV$ vertex 
 $\Gamma^{\mu_1\mu_2}_{XV_1V_2}(q_1,q_2;\lam)$ in
 eq.~(\ref{amp}), normalized by the scalar form factor
 $g_{XV_1V_2}^{}(q_1,q_2)$, are defined for $CP$-even and -odd Higgs bosons 
  ($H$ and $A$) and massive gravitons ($G$), respectively. For the
  polarization tensor $\eps^{\alpha\beta}(P,\lam)$ and the $GVV$
  vertex $\hat\Gamma^{\alpha\beta,\mu_1\mu_2}_{GVV}(q_1,q_2)$, see
  appendix~\ref{sec:spin2}.} 

For graviton resonances, we adopt the simplest RS 
model~\cite{Randall:1999ee,Davoudiasl:1999jd}, where only gravitons
can propagate into the extra dimension, and consider the first
excited mode of the Kaluza-Klein (KK) gravitons. 
The low-energy effective interactions with the SM fields are given by 
\begin{align}
 {\cal L}_G=-\frac{1}{\Lambda}\,T^{\mu\nu}G_{\mu\nu},
\label{L_G}
\end{align}
where $T^{\mu\nu}$ is the energy-momentum tensor of the SM fields (see,
{\it e.g.}, ref.~\cite{Hagiwara:2008jb} for the explicit forms) and
$G_{\mu\nu}$ is the spin-2 KK graviton.
The RS graviton excitations have the universal coupling strength of 
$1/\Lambda$ to the matter and gauge fields, {\it e.g.}
$g_{GVV}^{}=-1/\Lambda\ (V=W,Z,\gamma,g)$, where $\Lambda$ is the scale
parameter of the theory and can be a few TeV.  
The explicit forms of the polarization tensor for a spin-2 graviton and
the three-point $GVV$ vertices are given in appendix~\ref{sec:spin2}.\\ 

In table~\ref{Mh}, we present the reduced helicity amplitudes for the
$CP$-even/odd Higgs boson productions via off-shell vector-bosons, 
$\Mh_X{}_{\lambda_1\lambda_2}^{\lambda}
 (V_{\lambda_1}^*V_{\lambda_2}^*\to H/A_{\lambda}^{})$,
where
\begin{align}
   \big(\M^{X}_{V_1V_2}\big)^{\lam}_{\lam_1\lam_2}
 &=\eps_{\mu_1}(q_1,\lam_1)\,
   \eps_{\mu_2}(q_2,\lam_2)\,
   \Gamma^{\mu_1\mu_2}_{XV_1V_2}(q_1,q_2;\lam)^* \nn\\
 &=g_{XV_1V_2}^{}(q_1,q_2)\,\Mh_X{}_{\lambda_1\lambda_2}^{\lambda},
\label{rXVVamp}
\end{align}
in the VBF frame.
Since Higgs bosons are spin-0 particles ($\lambda=0$), we have the
helicity selection rule ($\lam_1=\lam_2$), and hence only three
amplitudes among the nine amplitudes which have the different helicity
combinations of the colliding vector-bosons can be nonzero. In the table, 
the amplitudes are expressed in terms of the Higgs boson
mass $M$ and magnitudes of the four-momentum squared of the vector
bosons $Q_1$ and $Q_2$. 
The amplitudes for the $CP$-even and $CP$-odd Higgs bosons via the
collisions of the transversely-polarized vector-bosons have
the relationships, respectively,
\begin{align}
 \Mh_H{}_{++}-\Mh_H{}_{--} &=0, \label{H++--}\\ 
 \Mh_A{}_{++}+\Mh_A{}_{--} &=0. \label{A++--}
\end{align}
Moreover, the pseudoscalar Higgs bosons ($A$) cannot be produced through
the longitudinal vector-bosons, $\Mh_A{}_{00}=0$, due to the $CP$-odd
property. 
We should notice that, in the case of $Q_1,Q_2\ll M$, 
the Higgs bosons via WBF are produced mostly through the 
longitudinally-polarized weak-bosons. On the other hand, the
loop-induced $CP$-even ($CP$-odd) Higgs bosons are produced mainly
(only) by the transversely-polarized vector-bosons, and the magnitude of
their amplitudes are almost equal, $|\Mh_H|\sim|\Mh_A|$, apart from the
overall coupling factors. 
These characters of the $XVV$ amplitudes, together with the phases of the
current amplitudes, play an important role to develop the distinctive
azimuthal angle correlations of the jets. 

\TABULAR[t]{|cc|ccc|}{\hline\\[-5mm]
 & &  \multicolumn{2}{c}{$CP$-even} & $CP$-odd \\
 $\lambda$ & $(\lam_1\lam_2)$ & $H$(WBF) & 
  $H$(loop-induced) & $A$ \\ \hline\\[-3mm]
 0 & $(\pm\pm)$ & $-1$ & $-\dfrac{1}{2}(M^2+Q_1^2+Q_2^2)$ 
   & $\mp\dfrac{i}{2}\sqrt{(M^2+Q_1^2+Q_2^2)^2-4Q_1^2Q_2^2}$ \\
 0 & (00)       & $\dfrac{M^2+Q_1^2+Q_2^2}{2Q_1Q_2}$ 
   & $Q_1Q_2$ & 0 \\[3mm]\hline}
{\label{Mh} The reduced helicity amplitudes for the $CP$-even/odd Higgs
 boson productions via off-shell vector-boson fusion,
 $\Mh_X{}^{\lam}_{\lam_1\lam_2}
  (V_{\lambda_1}^*V_{\lambda_2}^*\to H/A_{\lambda}^{})$, 
 in the VBF frame. $M$ is the Higgs boson mass, and $Q_1$ and $Q_2$
 are magnitudes of the four-momentum squared of the vector bosons.}

Similarly, table~\ref{MG} shows the reduced helicity amplitudes for the
massive-graviton productions via off-shell vector-bosons, 
$\Mh_X{}_{\lambda_1\lambda_2}^{\lambda}
 (V^*_{\lam_1}V^*_{\lam_2}\to G_{\lam}^{})$, 
in the VBF frame, where $\lam=\lam_1-\lam_2$ is the tensor helicity
along the colliding vector-boson axis (positive $z$-axis).
For the spin-2 particle productions, the amplitudes in all the helicity
combinations of the vector bosons exist, that is, the nine amplitudes are
registered in table~\ref{MG}. 
However, in the case of $Q_1,Q_2\ll M$, 
the $\lambda=\pm2$ states of the gravitons are dominantly produced for
the massless vector-boson collisions, when the vector bosons have
the opposite-sign transverse polarizations.
For the massive vector-boson case, on the other hand, the amplitudes of 
other three states ($\lambda=\pm1$ and $0$) are not negligible.

\TABULAR[t]{|cc|c|}{\hline\\[-5mm]
 $\lambda$ & $(\lam_1\lam_2)$ & $G$ \\ \hline\\[-3mm]
 $\pm 2$ & ($\pm\mp$) & $-(M^2+Q_1^2+Q_2^2+2m_V^2)$ \\[2mm]
 $\pm 1$ & $(\pm 0)$ 
         & $\dfrac{1}{\sqrt{2}MQ_2}
            \big[Q_2^2(M^2-Q_1^2+Q_2^2)-m_V^2(M^2+Q_1^2-Q_2^2)\big]$ \\[2.5mm]
 $\pm 1$ & $(0\mp)$ 
         & $\dfrac{1}{\sqrt{2}MQ_1}
            \big[Q_1^2(M^2+Q_1^2-Q_2^2)-m_V^2(M^2-Q_1^2+Q_2^2)\big]$  \\[2.5mm]
 0 & $(\pm\pm)$ & $\dfrac{1}{\sqrt{6}M^2}
     \big[(Q_1^2-Q_2^2)^2+M^2(Q_1^2+Q_2^2-2m_V^2)\big]$ \\[2.5mm]
 0 & $(00)$ & \hspace*{-2.5cm}
     $-\dfrac{1}{\sqrt{6}Q_1Q_2}\Big[4Q_1^2Q_2^2
      +2m_V^2(M^2+Q_1^2+Q_2^2)$ \\ [-1.5mm]
  & & \hspace*{3cm}$-\dfrac{m_V^2}{M^2}\{(M^2+Q_1^2+Q_2^2)^2-4Q_1^2Q_2^2\}\Big]$ \\[3mm]\hline} 
{\label{MG} The same as table~\ref{Mh}, but for the massive-graviton
 productions, $V^*_{\lam_1}V^*_{\lam_2}\to G_{\lam}^{}$, where $M$ and
 $m_V$ is the graviton and the vector-boson mass, respectively.} 

The $X\to VV$ decay helicity amplitudes in eq.~(\ref{dVBFamp}), 
$\big(\M'^{\,X}_{\ V'_1V'_2}\big)^{\lam'}_{\lam'_1\lam'_2}$, are
obtained from the above $VV\to X$ fusion amplitudes (tables~\ref{Mh} and
\ref{MG}) by making the 
replacements $Q_i^2\to -{Q'_i}^2$, where $\lam'_1$ and $\lam'_2$ are the
helicities of the decaying vector-bosons and $\lam'=\lam'_1-\lam'_2$
is the $X$ helicity along the momentum direction of the vector bosons
($z'$-axis in fig.~\ref{fig:frame}).
We note that for the non-scalar particle decays the dependence of the
angle $\Theta$ between the 
initial polarization $\lambda$ along the $z$-axis and the final
polarization $\lambda'$ along the $z'$-axis is dictated
by a $d$ function and factorized in our convention; see eq.~(\ref{pdamp}). 

At this point it is worth noting that the amplitudes of the Higgs boson
decays into transversely-polarized weak-bosons  ($\Mh'_H{}_{\pm\pm}$)
are comparable to that into longitudinal ones ($\Mh'_H{}_{00}$) near the
$VV$ threshold, while the longitudinal amplitudes are dominant for large
Higgs boson masses. This is because one or both of the weak bosons can
become on-shell and the typical mass scale of the decaying weak-bosons
are $Q_i\sim m_V$. The helicity amplitudes of the Higgs boson decays
into the on-shell weak-bosons are given by
\begin{align}
  \Mh'_H{}_{\pm\pm}=-1,\quad \Mh'_H{}_{00}=\frac{M^2}{2m_V^2}-1
 =\gamma'^2(1+\beta'^2),
\label{HVVdecay}
\end{align}
where $\beta'=\sqrt{1-4m_V^2/M^2}$ and $\gamma'=1/\sqrt{1-\beta'^2}$ are
the velocity and the Lorentz-boost factor of the weak bosons,
respectively.%
\footnote{See ref.~\cite{Choi:2002jk} for the amplitudes of
the Higgs boson decays into a pair of virtual and real weak-bosons with
the most general $HV^*V$ vertices.} 

It may be also useful to present the helicity amplitudes for the
graviton decays into a on-shell vector-boson pair; see Table~\ref{MGVV}, 
which can be reduced from the
$VV\to G$ amplitudes in table~\ref{MG} with the replacements
$Q_i^2\to-m_V^2$. 
In the heavy graviton-mass limit, or in the $\beta'=1$ limit, only three
(two) amplitudes among the nine survive for the decays into massive
(massless) vector-bosons. 

\TABULAR{|cc|c|}{\hline\\[-4.5mm]
 $\lam'$ & $(\lam'_1\lam'_2)$ & $G$ 
 \\[0.5mm] \hline\\[-4mm]
 $\pm 2$ & $(\pm\mp)$ & $-M^2$ \\[1mm]
 $\pm 1$ & $(\pm 0),(0\mp)$
         & $\frac{1}{\sqrt{2}}\sqrt{1-\beta'^2}\,M^2$ \\[1mm]
 $    0$ & $(\pm\pm)/(00)$ & $-\frac{1}{\sqrt{6}}(1-\beta'^2)\,M^2\,\big/
      -\frac{1}{\sqrt{6}}(2-\beta'^2)\,M^2$ \\[2mm]\hline}
{\label{MGVV} The reduced helicity amplitudes for the massive-graviton
 decays into a on-shell vector-boson pair, 
$\Mh'_X{}^{\lam'}_{\lam'_1\lam'_2}(G_{\lam'}^{}\to V_{\lam'_1}V_{\lam'_2})$, 
 in the graviton rest frame.}

\section{Azimuthal angle correlations}\label{sec:az}

We are now ready to present the helicity amplitudes explicitly for the
$X$ productions with 2 jets via VBF, eq.~(\ref{pamp}), and the $X$
decays into 4 jets via a vector-boson pair, eq.~(\ref{damp}).
In this section, we demonstrate that the VBF amplitudes dominate the
exact matrix elements by appropriate selection cuts to the final
state~\cite{Hagiwara:2009zz}, and then show that nontrivial 
azimuthal angle correlations of the jets in the production are
manifestly expressed as the quantum interference among different
helicity states of the intermediate vector-bosons. We also discuss the
$X$ decay angular correlations.

\subsection{The VBF amplitudes vs. the full amplitudes}
\label{sec:VBFvsfull} 

Although we have considered only the VBF diagrams in fig.~\ref{fig:vbf},
there are other crossing-related diagrams which have to be taken into
account. Representative Feynman diagrams for each subprocess, including
the VBF diagrams, are shown in fig.~\ref{fig:pp_jjX}. 

\FIGURE[t]{
 \epsfig{file=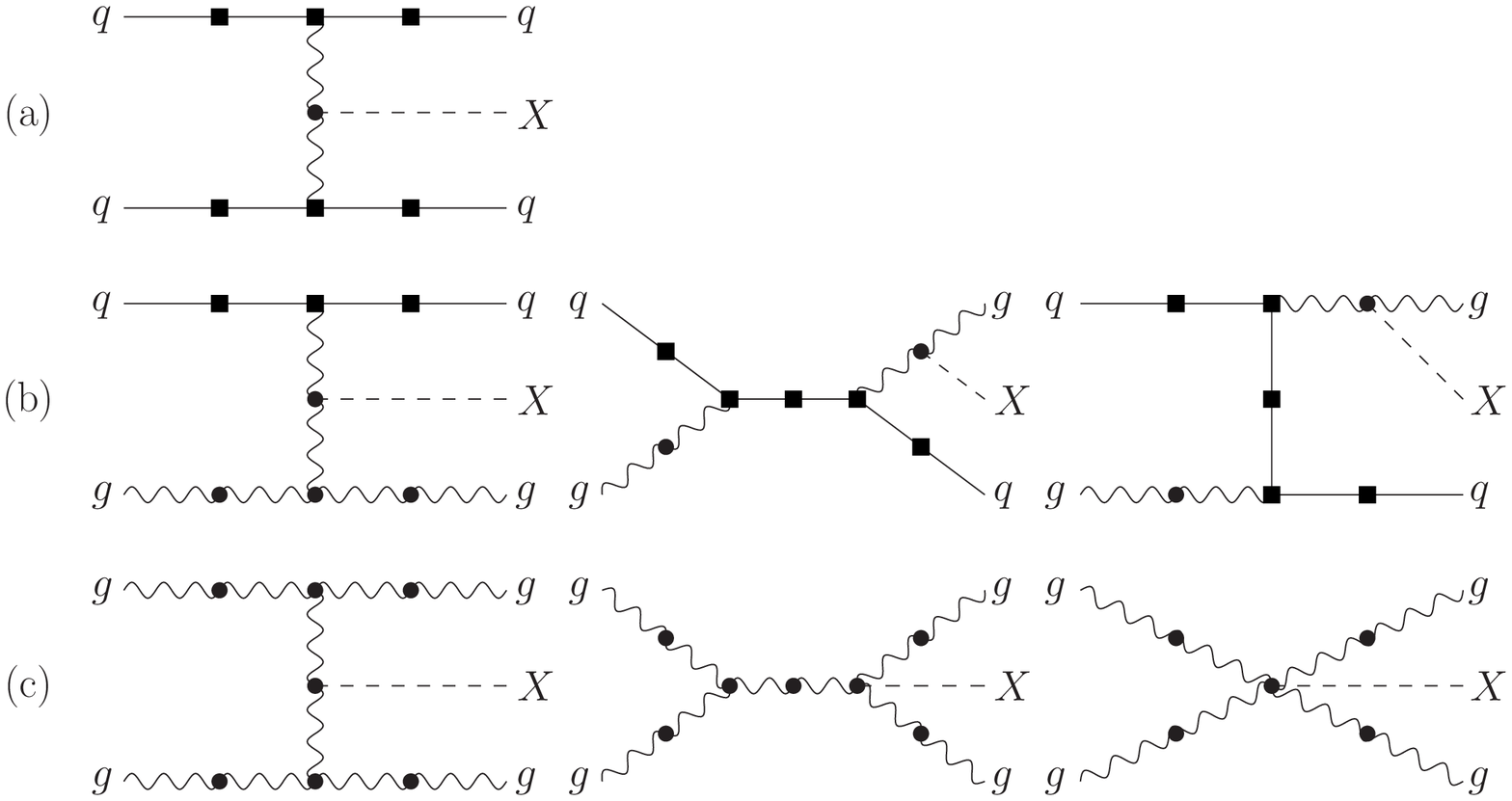,width=1\textwidth,clip}
 \caption{Representative Feynman diagrams for the subprocesses, (a)
 $qq\to qqX$, (b) $qg\to qgX$, (c) $gg\to ggX$, which contribute to the
 $X$ + 2-jet production at hadron colliders, $pp\to jjX$. The Higgs
 bosons are emitted from each of the circle points in the diagrams,
 while the KK gravitons are emitted from each of the circle and square
 points.  
 \label{fig:pp_jjX}}}

The Higgs bosons are emitted from each of the circle points in the
diagrams. 
(a)~$qq\to qqH/A$: The Higgs boson is radiated off the weak-boson
propagator, or the gluon (photon) propagator through a top-quark
triangle-loop. There is only a $t$-channel VBF diagram in the case of
the different-flavor initial state, while an additional $s$-channel
diagram or a $u$-channel diagram exists in the identical-flavor case.   
(b)~$qg\to qgH/A$: In addition to the $H/Agg$ coupling along the gluon
lines, the process receives the $H/Aggg$ box-loop contribution on the
gluon three-point vertex. There are in total 8 diagrams
in the large-$m_t$ limit,%
\footnote{Away from the heavy-top limit, the number of diagrams
increases because of the ordering of the gluon momenta along the
top-quark loop~\cite{DelDuca:2001eu}.}
in which the top-quark loop is replaced by the effective coupling. 
(c)~$gg\to ggH/A$: Besides the three corresponding $gg\to gg$ processes in 
fig.~\ref{fig:pp_jjX}(c), the $u$-channel diagram exists. There are 26
diagrams in the heavy-top limit, including one diagram with the
effective $Hgggg$ vertex induced via a pentagon loop.%
\footnote{A $CP$-odd Higgs boson does not have the $Agggg$ vertex due
to the anti-symmetric nature of the coupling~\cite{Kauffman:1998yg}.}

The KK gravitons are emitted from both the circle
and the square points in the diagrams in fig.~\ref{fig:pp_jjX}, due to
their universal couplings to the matter and gauge fields.
Therefore, the graviton productions have many more diagrams even in
quark-quark scatterings (fig.~\ref{fig:pp_jjX}(a)).

The point which must be investigated here is whether the VBF amplitudes can
dominate the exact matrix elements, in which all the possible diagrams
contribute. Our key observation is that this happens when we select
those events which satisfy the characteristic kinematical
structure of the VBF processes. Due to the $t$-channel
propagators of the vector bosons in the VBF amplitudes in 
eq.~(\ref{pamp}), the $Xjj$ events via VBF are dominantly produced when both
$Q_1\,\big(=\sqrt{-q_1^2}\,\big)$ and $Q_2\,\big(=\sqrt{-q_2^2}\,\big)$
are small. 
In other words, the intermediate vector-bosons in the VBF processes tend
to carry only a small fraction of the initial parton energies.
For small $Q_1$ and $Q_2$, the initial partons scatter to far
forward and far backward, and the heavy particle $X$ is produced
centrally. Therefore, the two jets have the large rapidity gap, which is
often used as the so-called WBF cut to enhance the WBF Higgs productions. 
It should be stressed that this kinematical feature is not particular
to the WBF production processes, but the 
QCD productions via the $t$-channel GF processes in $qq$, $qg$ and $gg$
collisions also have the similar kinematical structure. 
This suggests that some kinematical cuts, such as a large rapidity
separation between two jets, may select the VBF diagrams dominantly
among all the possible diagrams.\\

Let us demonstrate numerically that the dijet large rapidity separation 
is an effective kinematical cut to select the VBF amplitudes among the
full amplitudes. 
As the minimal selection cuts on the final-state partons,
we impose the following kinematical constraints for the LHC, required by
the detector and jet algorithms: 
\begin{align}
 p_{T_j}>20\ {\rm GeV},\quad |\eta_j|<5,\quad 
 R_{jj}=\sqrt{\Delta\eta_{jj}^2+\Delta\phi_{jj}^2}>0.6,
\label{mincut}
\end{align}
where $p_{T_j}$ and $\eta_j$ are the transverse momentum and the
pseudorapidity of a final-state parton, respectively, and $R_{jj}$
describes the separation of the two partons in the plane of the
pseudorapidity and the azimuthal angle. 
Moreover, in order to select the VBF contributions, the two tagging jets
are required to reside in opposite detector hemispheres and to be well
separated in rapidity,
\begin{align}
 \eta_{j_1}>0>\eta_{j_2},\quad
 \Delta\eta_{jj}=\eta_{j_1}-\eta_{j_2}>\Delta\eta_{jj\rm min}.
\label{vbfcut}
\end{align}
Varying the value of $\Delta\eta_{jj\rm min}$, we study the fraction of
the VBF contributions to the cross section with the exact matrix elements. 
The analyses are done at the parton level with tree-level matrix elements.  
The exact matrix elements for the Higgs boson productions are calculated
by {\tt HELAS} subroutines~\cite{Murayama:1992gi}, generated
by the {\tt HEFT} (Higgs effective field theory) model in 
{\tt MadGraph/MadEvent} ({\tt MG/ME}) {\tt v4}~\cite{Alwall:2007st}. 
For the massive-graviton productions, the relevant 
{\tt HELAS} subroutines for massive spin-2 particles and its
interactions based on the effective Lagrangian of eq.~(\ref{L_G}) have 
also been implemented into {\tt MG/ME}~\cite{Hagiwara:2008jb}. 
Numerical integrations are done with the help of the Monte Carlo
integration program {\tt BASES}~\cite{Kawabata:1995th}.
Throughout our numerical study, we employ the CTEQ6L1 parton distribution 
functions~\cite{Pumplin:2002vw} with the factorization scale chosen as
the geometric mean of the jet transverse momenta 
$\mu_f=\sqrt{p_{T_{j_1}}p_{T_{j_2}}}$, and fix the QCD coupling at
$\alpha_s=\alpha_s^{\rm LO}(m_Z)=0.13$.
Unless specified, we set the heavy particle mass at $M=600$ GeV and the
scale of the RS model at $\Lambda=4$ TeV, which
corresponds to the current lower bound for the mass of the first KK
mode of massive gravitons~\cite{Davoudiasl:1999jd,Abazov:2005pi}.
Note that we use the constant value for the Higgs effective couplings in
spite of $M_{H,A}>2m_t$, because the energy dependence of the effective
couplings are almost canceled out when the ratio of the cross
sections is considered, and because the azimuthal angle distributions are
insensitive to the form factor effects~\cite{Figy:2004pt}.

\TABULAR[t]{|c||ccc|}
 {\hline 
   $\sigma_{\rm VBF}/\sigma_{\rm exact}$
 & $\Delta\eta_{jj}>3$ & $\Delta\eta_{jj}>4$ & $\Delta\eta_{jj}>5$ 
 \\ \hline
  $qq\to qqH/A/G$ & 1.00/1.00/1.58 & 1.00/1.00/1.43 & 1.00/1.00/1.25 \\
  $qg\to qgH/A/G$ & 1.07/1.05/1.30 & 1.04/1.03/1.18 & 1.02/1.02/1.11 \\
  $gg\to ggH/A/G$ & 1.07/1.06/1.16 & 1.04/1.04/1.11 & 1.02/1.02/1.07 \\ 
 \hline} 
{\label{Xsec} Ratio of the VBF contribution to the cross section with
   the exact matrix elements, $\sigma_{\rm VBF}/\sigma_{\rm exact}$,
 for each subprocess at the LHC, after imposing the inclusive 
 cuts~(\ref{mincut}) and the VBF cuts~(\ref{vbfcut}) with 
  $\Delta\eta_{jj\rm min}=$ 3, 4 and 5.} 

In table~\ref{Xsec}, we show the ratio of the VBF contribution 
($\sigma_{\rm VBF}$) to the cross section with the exact matrix elements
($\sigma_{\rm exact}$)
for the nine subprocesses at the LHC, after imposing
the inclusive cuts~(\ref{mincut}) and the VBF selection cuts~(\ref{vbfcut}) 
with $\Delta\eta_{jj\rm min}=$ 3, 4 and 5. 
For all the subprocesses, as the rapidity separation increases, the VBF
contributions tend to dominate the exact matrix elements and the ratios
approach unity.
In the first row, the $CP$-even and -odd Higgs production processes in
$qq$ collisions, 
where we consider $ud$ collisions for simplicity, have only the VBF 
diagrams, and hence $\sigma_{\rm VBF}/\sigma_{\rm exact}=1$.%
\footnote{Even for the collisions of identical-flavor quarks, the
 $s$- or $u$-channel contribution is
 negligible when the VBF cuts are applied~\cite{Ciccolini:2007jr}. 
 In addition, the interferences between the electroweak and QCD
 contributions are very 
 small~\cite{Andersen:2006ag,Andersen:2007mp,Bredenstein:2008tm}.}
Although the cross sections in $qg$ and $gg$ collisions have non-VBF
diagrams, their contributions are rather small after the VBF cuts.
On the other hand, for the massive-graviton production, the gravitons
are emitted also from the quark lines, as shown in
figs.~\ref{fig:pp_jjX} (a) and (b), and the ratio deviates significantly
from unity especially in $qq$ collisions. Furthermore, their
contributions do not diminish swiftly when the rapidity separation cut
is increased from 3 to 5. 
We note here that the electroweak contributions to the $qq\to qqG$ process
represent a small correction, which is below 1\%, even when the VBF cuts
are imposed~\cite{Hagiwara:2008iv}.

\TABULAR[t]{|c||ccc|}
 {\hline 
   $\sigma_{\rm VBF}/\sigma_{\rm exact}$
 & $\Delta\eta_{jj}>3$ & $\Delta\eta_{jj}>4$ & $\Delta\eta_{jj}>5$ 
 \\ \hline
  $qq\to qqH/A/G$ & 1.00/1.00/1.02 & 1.00/1.00/1.02 & 1.00/1.00/1.02 \\
  $qg\to qgH/A/G$ & 1.04/1.04/1.07 & 1.03/1.03/1.06 & 1.02/1.02/1.04 \\
  $gg\to ggH/A/G$ & 1.05/1.05/1.09 & 1.04/1.04/1.07 & 1.02/1.02/1.05 \\
 \hline} 
{\label{Xsec_ptcut} The same as table~\ref{Xsec}, but imposing the additional
   $p_{T_j}$ cut of eq.~(\ref{ptcut}).}

In case of the massive-graviton production, the non-VBF contributions
are significant even after the VBF
selection cuts with $\Delta\eta_{jj\rm min}=5$. We therefore examine the
impact of an additional cut on the transverse momenta of the tagging
jets, 
\begin{align}
 p_{T_j}<100\ {\rm GeV}.
\label{ptcut}
\end{align}
Table~\ref{Xsec_ptcut} shows the same ratio of the VBF contribution to
the exact cross section when this additional cut is imposed. We find
that the above $p_{T_j}$ cut works effectively to suppress contributions
from the non-VBF diagrams, especially the diagrams which emit the
graviton from the quark lines.
This is because the quarks that emit a graviton tends to have high
transverse momenta. 

Summing up, from table~\ref{Xsec}, the large rapidity separation may
guarantee the validity of our VBF analyses not only for the WBF Higgs
productions but also for the GF processes. Moreover, from
table~\ref{Xsec_ptcut}, the $p_{T_j}$ slicing cut, 
\begin{align}
 20\ {\rm GeV}<p_{T_j}<100\ {\rm GeV},
\label{ptslicingcut}
\end{align}
from eqs.~(\ref{mincut}) and (\ref{ptcut}) is effective in selecting the
GF contribution to the graviton production processes. 
It should be noticed that stringent cuts increase the VBF contributions
but reduce the primary event number.  

\TABULAR[b]{|c||ccc|ccc|}
 {\hline
 & \multicolumn{3}{c|}{$20\ {\rm GeV}<p_{T_j}$} &
   \multicolumn{3}{c|}{$20\ {\rm GeV}<p_{T_j}<100\ {\rm GeV}$}  \\
   $\sigma_{\rm exact}$\,[pb]
 & $\Delta\eta_{jj}>3$ & $\Delta\eta_{jj}>4$ & $\Delta\eta_{jj}>5$ 
 & $\Delta\eta_{jj}>3$ & $\Delta\eta_{jj}>4$ & $\Delta\eta_{jj}>5$ \\ 
 \hline
  $qq\to qqG$ &  2.0 & 1.6 & 1.1
              &  0.7 & 0.6 & 0.5 \\
  $qg\to qgG$ & 13.2 & 8.8 & 4.7 
              &  7.2 & 5.5 & 3.3 \\
  $gg\to ggG$ & 15.9 & 8.0 & 3.2
              & 11.7 & 6.2 & 2.7 \\
 \hline}
 {\label{Xsec_G}The total cross sections with
 the exact matrix elements for each subprocess of the massive-graviton 
 plus dijet events at the LHC, 
 after imposing the inclusive 
 cuts~(\ref{mincut}) and the VBF cuts~(\ref{vbfcut}) without and with 
 the additional $p_{T_j}$ cut of eq.~(\ref{ptcut}), for $M_G=600$ GeV 
 and $\Lambda=4$ TeV.}

As a reference, we present the total cross sections with the exact matrix 
elements for the massive-graviton productions in table~\ref{Xsec_G}, 
where the inclusive 
 cuts~(\ref{mincut}) and the VBF cuts~(\ref{vbfcut}) without and with 
 the $p_{T_j}$ slicing cut (\ref{ptslicingcut}) are imposed.
For the cross section of the Higgs boson productions, see fig.~4 in
ref.~\cite{DelDuca:2001eu}. 
In the following analyses, we take $\Delta\eta_{jj\rm min}=4$ 
in the VBF cuts (\ref{vbfcut}) for the Higgs and graviton productions, 
and further apply the $p_{T_j}$ slicing cut~(\ref{ptslicingcut}) for the 
graviton productions.

\subsection{Correlations in the production with two associated jets}

We are now able to discuss the angular correlations of the two
accompanying jets in the $X$ productions, using our analytical VBF
amplitudes. 

The production and decay density matrices in eq.~(\ref{M2}) have all
information on the 
angular correlations between the jets in the productions and the decays
of the heavy particle $X$.
In this section, we consider the $n=2$ case in (\ref{M2}) to investigate
the angular correlation between the two jets in the $Xjj$ productions,
while, to simplify the decay part, we consider the $n'=2$ case and 
fix the $X$ polarization along the 2-body decay axis ($z'$-axis),
$\lambda'=\sig'_1-\sig'_2=\pm 2$, $\pm 1$, or $0$.
We note that, in practice, we can project out $\sig'_1$ and $\sig'_2$ by properly
weighting the final states of $a'_1$ and $a'_2$ decays in the $X\to
a'_1a'_2$ decays; 
{\it e.g.} for the $X\to W^+W^-$ decays we can project out all the five
cases, $\lam'=\pm 2$, $\pm 1$, and $0$, while the $X\to\gamma\gamma$ or
$gg$ decays give only the sum of $\lam'=+2$ and $-2$.
Moreover, the $X\to\tau^+\tau^-$ decay process can project out $\lambda'=+1$ and
$-1$ cases, while for the $e$ or $\mu$ case we cannot distinguish
these two. 
In the following we take $\lam'=\bar\lam'$ and suppress the decay
density matrix in eq.~(\ref{M2}).

For the VBF processes (\ref{ff_ffx}), the production density matrix is given in 
eq.~(\ref{PtoPtensor}) in terms of the production tensor
$\P^{\lam_1\lam_2}_{\bar\lam_1\bar\lam_2}$ of eq.~(\ref{ptensor}).
The tensor has
81 independent jet angular distributions in terms of the polar
($\theta_{1,2}$) and azimuthal ($\phi_{1,2}$) angles of the two tagging jets.
When we isolate the azimuthal angle dependence in eq.~(\ref{M2}), there are
25 distributions (including one constant piece) as  
\begin{align}
  \P^{\lam_1\lam_2}_{\bar\lam_1\bar\lam_2}
  d^J_{\lam,\lam'}d^J_{\bar\lam,\lam'}
 = F_1^{}+\big\{2\,{\Re}e&\big[F_2^{}\cos\phi_1+F_3^{}\cos\phi_2
  +F_4^{}\cos2\phi_1+F_5^{}\cos2\phi_2 \nn\\
 &+F_6^{\pm}\cos(\phi_1\pm\phi_2)
  +F_7^{\pm}\cos(2\phi_1\pm\phi_2)+F_8^{\pm}\cos(\phi_1\pm2\phi_2) \nn\\
 &+F_9^{\pm}\cos2(\phi_1\pm\phi_2)\big]
  +({\Re}e\to{\Im}m,\,\cos\to\sin)\big\}.
\label{general_azdis}
\end{align}
Here, and in the following, summation over repeated indices 
$(\lam_1,\lam_2,\bar\lam_1,\bar\lam_2)=\pm,0$ 
is implied, and 
a shorthand notation such as $F_6^{\pm}\cos(\phi_1\pm\phi_2)$ for 
$F_6^{+}\cos(\phi_1+\phi_2)+F_6^{-}\cos(\phi_1-\phi_2)$ is used. 
The coefficients $F_i^{(\pm)}$ are the functions of the kinematical
variables except the azimuthal angles $\phi_{1,2}$.
For the productions of spin-full heavy particles, they also
depend on the decay angle $\Theta$ which comes from the product of
two $d$ functions.
For the spin-0 particle case, only the five terms in
eq.~(\ref{general_azdis}) survive due to the
helicity selection $\lam_1=\lam_2$, which will be discussed in 
the next subsection~\ref{sec:azcor_H}.
All the {\it sine} terms vanish when $CP$ is conserved and when the
absorptive part of the amplitudes are neglected, {\it e.g.}, in the
tree-level approximation.
It should be noted that the azimuthal angle variables,
$\phi_{1}$ and $\phi_{2}$, are individually defined 
in the VBF frame by the scattering plane of the subprocess, 
$V^*_1V^*_2\to X\to a'_1a'_2$; see also
fig.~\ref{fig:frame}. 

Because the phases of the quark and gluon current amplitudes are the
same for each helicity combination (see tables~\ref{Jq}(top) and
\ref{Jg}(top)), and because the phase of the product of the two currents 
for $\sigma_1=\sigma_3$ and
$\sigma_2=\sigma_4$ is (see eq.~(\ref{J2in_con}))
\begin{align}
 \Jh_1{}^{\lam_1}_{\sig_1,\sig_3=\sig_1}
 \Jh_2{}^{\lam_2}_{\sig_2,\sig_4=\sig_2}\propto
 e^{-i(\lam_1\phi_1-\lam_2\phi_2)},
\label{J1J2phase}
\end{align}
the coefficients $F_{1-9}^{(\pm)}$ for $qq$, $qg$ and $gg$ collisions
are expressed in terms of the production tensors
$\P^{\lam_1\lam_2}_{\bar\lam_1\bar\lam_2}$
and two $d$ functions as
\begin{align}
 F_1^{}&=\P^{\lam_1\lam_2}_{\lam_1\lam_2}
  d^J_{\lam_1-\lam_2,\lam'}d^J_{\lam_1-\lam_2,\lam'}, \nn\\
 F_2^{}&= \P^{+\lam_2}_{0\lam_2}
  d^J_{1-\lam_2,\lam'}d^J_{-\lam_2,\lam'}
         +\P^{0\lam_2}_{-\lam_2}
  d^J_{-\lam_2,\lam'}d^J_{-1-\lam_2,\lam'}, \nn\\
 F_3^{}&= \P^{\lam_10}_{\lam_1+}
  d^J_{\lam_1,\lam'}d^J_{\lam_1-1,\lam'}
         +\P^{\lam_1-}_{\lam_10}
  d^J_{\lam_1+1,\lam'}d^J_{\lam_1,\lam'}, \nn\\
 F_4^{}&=\P^{+\lam_2}_{-\lam_2}
  d^J_{1-\lam_2,\lam'}d^J_{-1-\lam_2,\lam'}, \nn\\
 F_5^{}&=\P^{\lam_1-}_{\lam_1+}
  d^J_{\lam_1+1,\lam'}d^J_{\lam_1-1,\lam'}, \nn\\
 F_6^{\pm}&= \P^{+0}_{0\pm}
  d^J_{1,\lam'}d^J_{-1/1,\lam'}
            +\P^{+\mp}_{00}
  d^J_{2/0,\lam'}d^J_{0,\lam'}
            +\P^{00}_{-\pm}
  d^J_{0,\lam'}d^J_{-2/0,\lam'}
            +\P^{0\mp}_{-0}
  d^J_{1/-1,\lam'}d^J_{-1,\lam'}, \nn\\
 F_7^{\pm}&= \P^{+0}_{-\pm}
  d^J_{1,\lam'}d^J_{-2/0,\lam'}
            +\P^{+\mp}_{-0}
  d^J_{2/0,\lam'}d^J_{-1,\lam'}, \nn\\
 F_8^{\pm}&= \P^{+\mp}_{0\pm}
  d^J_{2/0,\lam'}d^J_{-1/1,\lam'}
            +\P^{0\mp}_{-\pm}
  d^J_{1/-1,\lam'}d^J_{-2/0,\lam'}, \nn\\
 F_9^{\pm}&=\P^{+\mp}_{-\pm}
  d^J_{2/0,\lam'}d^J_{-2/0,\lam'}. 
\label{F1to9}
\end{align}
This relations can be also applied to 
the processes with the helicity-flip currents, namely 
the cases for $\sigma_1=-\sigma_3$ and/or $\sigma_2=-\sigma_4$, 
which appear only in the gluon currents, although they
have the different phase from those
with the conserved currents in eq.~(\ref{J1J2phase}). 
The process that one of the currents is helicity-flip
(the case in $\sig_1=-\sig_3$ or $\sig_2=-\sig_4$) leads nontrivial
azimuthal distributions from
$F_2$ through $F_5$, while the case that the
both currents are helicity-flip gives rise to only the constant piece,
$F_1$. 
This is because the
helicity-flip gluon splitting amplitudes emit an off-shell gluon of
definite helicity; see table~\ref{Jg}.
On the other hand, the helicity-conserved processes ($\sigma_1=\sigma_3$
and $\sigma_2=\sigma_4$) can contribute to all the terms, as
can be seen from table~\ref{Jq} for the quark 
currents and table~\ref{Jg} for the gluon currents.
We note that the magnitude of the correlations is determined
by the relative ratio to the constant term $F_1$.

It should be stressed here that 
in eq.~(\ref{F1to9}) the $\Theta$-dependent
azimuthal angle correlations are manifestly expressed by quantum 
interference among different helicity states of the
intermediate vector-bosons.
We also notice that the above formulae can be applied to any
spin-$J$ particle productions through the VBF processes, although
massive spin-0 and -2 particles are considered in this article.

\subsubsection{Higgs boson productions}\label{sec:azcor_H}

For the scalar particle productions,
only the three off-shell VBF amplitudes, $\Mh_X{}_{++}$,
$\Mh_X{}_{00}$ and $\Mh_X{}_{--}$, in which the 
colliding vector-bosons have the same helicities ($\lambda_1=\lambda_2$),
can contribute to the production amplitude~(\ref{pamp}), and 
there is no $\Theta$ dependence, namely 
$d^{J=0}_{\lam,\lam'}(\Theta)=1$ in eq.~(\ref{general_azdis}) (and in
eq.~(\ref{F1to9})). Therefore, 
the azimuthal angle correlation~(\ref{general_azdis}) is reduced to
\begin{align}
  \P^{\lam_1\lam_2}_{\bar\lam_1\bar\lam_2}
 = F_1^{}+\big\{2\,{\Re}e\big[F_6^{-}\cos\Delta\phi_{12}
  +F_9^{-}\cos2\Delta\phi_{12}\big]
  +({\Re}e\to{\Im}m,\,\cos\to\sin)\big\}
\label{dphidep0}
\end{align}
with $\Delta\phi_{12}\equiv\phi_1-\phi_2$, which is the azimuthal angle
separation of the two tagging jets.
As mentioned above, the azimuthal dependence is manifestly expressed by
the quantum interference terms among different helicity states of the
intermediate vector-bosons; 
the $F_6^{-}\,(=\P^{++}_{00}+\P^{00}_{--})$ term is induced through the
interference between the production with the longitudinal and transverse
polarization states of the vector bosons,
while the $F_9^{-}\,(=\P^{++}_{--})$ term is caused by the interference
between the two transverse polarization states.

Now we can observe clearly the origin of the nontrivial azimuthal angle
correlations for the $CP$-even and $CP$-odd Higgs bosons ($H$ and $A$),
predicted in  
refs.~\cite{Plehn:2001nj,DelDuca:2001eu,Hankele:2006ja,Klamke:2007cu}.
As we mentioned in section~\ref{sec:vbfamp}, in the case of 
$Q_1,Q_2\ll M$, where the VBF contributions are dominant, the Higgs
bosons with a $g^{\mu\nu}$-type coupling are produced mostly through the
longitudinally-polarized vector-bosons. Therefore, the $\Mh_H{}_{00}$
amplitude dominates the total amplitudes, and hence there 
is little interference terms in eq.~(\ref{dphidep0}). This is why the WBF
processes give the flat azimuthal angle correlation,
\begin{align}
 d\hat\sigma_{H\rm (WBF)} \sim F_1.
\label{dphiWBF}
\end{align}

For the loop-induced GF Higgs boson couplings, on the other hand, they
are mainly produced by the transversely-polarized vector-bosons, namely
the $\Mh_{H/A}{}_{++}$ and $\Mh_{H/A}{}_{--}$ amplitudes have the
dominant contribution. Therefore, by the relations in (\ref{H++--}) and
(\ref{A++--}), the azimuthal distributions are 
\begin{align}
 d\hat\sigma_{H/A\rm (GF)} \sim 
   \big[F_1\pm 2\,|F_9^-|\cos2\Delta\phi_{12}\big],
\label{dphiGF}
\end{align}
where the $+/-$ sign is for $CP$-even/odd Higgs bosons.
One can clearly see that the azimuthal distribution is strongly
suppressed (enhanced) around $\Delta\phi_{12}=\pi/2$ for the GF
$CP$-even (-odd) Higgs boson productions. 
We note that the $F_6^-$ term in eq.~(\ref{dphidep0}) is exactly zero
not only for the $CP$-odd Higgs boson production but also for the GF
$CP$-even Higgs boson production since it measures the $P$-odd amplitude.  

From the relations in (\ref{F1to9}) and by
using the explicit forms of the amplitudes in tables~\ref{Jq}(top), 
\ref{Jg}(top) and \ref{Mh}, the coefficient functions $F_1$ and $F_9^-$ in 
eq.~(\ref{dphiGF}) for each subprocess, $qq$, $qg$ and 
$gg$ scatterings, are given by
\begin{subequations}
\begin{align}
 \hat F_1[qq]&=(1+\cos^2\theta_1)(1+\cos^2\theta_2), \\
 \hat F_1[qg]&=\dfrac{1}{\sin^2\theta_2}
                   (1+\cos^2\theta_1)(1+3\cos^2\theta_2)^2 
                  \quad {\rm or}\quad (1\leftrightarrow 2), \\
 \hat F_1[gg]&=\dfrac{1}{\sin^2\theta_1\sin^2\theta_2}
                   (1+3\cos^2\theta_1)^2(1+3\cos^2\theta_2)^2,
\end{align}
\label{F1}%
\end{subequations}
and
\begin{align}
 \hat F_9^-[qq/qg/gg]=\pm \frac{1}{2}(1-\cos^2\theta_1)(1-\cos^2\theta_2).
\label{F9-}
\end{align}
Here we take the $Q_1,Q_2\ll M$ limit for the $VV\to X$ amplitudes,
where the only surviving amplitudes are 
$\big|\Mh_{H/A}{}_{\pm\pm}\big|=\frac{1}{2}M^2$ (see table~\ref{Mh}), with the
common overall factor 
\begin{align}
  F_{i}^{(\pm)}
 =\frac{g_s^4g_{H/Agg}^2M^4Q_1^2Q_2^2}{2\cos^2\theta_1\cos^2\theta_2}\,
  \hat F_{i}^{(\pm)}.
\label{sigmahat}
\end{align}
Note that we suppress the color factors, which are relevant to the ratio
of the $qq$, $qg$ and $gg$ contributions in realistic simulations.   
It is remarkable that the interference term $F_9^-$, which receives the
contribution only from the helicity-conserved amplitudes
($\sigma_1=\sigma_3$ and $\sigma_2=\sigma_4$), is same for all $qq$,
$qg$ and $gg$ collision processes. Meanwhile   
$F_1$ has the different contributions from the quark 
currents and the gluon currents, which includes the helicity-flip
contributions ($\sigma_1=-\sigma_3$ and/or $\sigma_2=-\sigma_4$), 
and can be larger as the process
involves the gluon currents. 
These indicate that the gluon currents reduce the interference effect.
It may be worth presenting the above functions in terms of the $z_{1,2}$
variables in the collinear limit ($\beta_{1,2}\to 1$) in
eq.~(\ref{cos1z1t}): 
\begin{subequations}
\begin{align}
 \hat F_1[qq]&=4(2-2z_1+z_1^2)(2-2z_2+z_2^2), \\
 \hat F_1[qg]&=\dfrac{8}{(1-z_2)}
                   (2-2z_1+z_1^2)(1-z_2+z_2^2)^2
                  \quad {\rm or}\quad (1\leftrightarrow 2), \\
 \hat F_1[gg]&=\dfrac{16}{(1-z_1)(1-z_2)}
                   (1-z_1+z_1^2)^2(1-z_2+z_2^2)^2,
\end{align}
\end{subequations}
and
\begin{align}
 \hat F_9^-[qq/qg/gg]=\pm 8(1-z_1)(1-z_2),
\end{align}
with
\begin{align}
 F_{i}^{(\pm)}=
 \frac{g_s^4g_{H/Agg}^2M^4Q_1^2Q_2^2}{2z_1^2z_2^2}\,
 \hat F_{i}^{(\pm)}.
\end{align}

\FIGURE[t]{
 \epsfig{file=dphi.eps,width=1\textwidth,clip}
 \caption{Normalized azimuthal correlations
 $\Delta\phi_{12}$ (mod $2\pi$) between the two tagging jets in the
 $pp\to jjX$ process at the LHC, where the selection
 cuts (\ref{mincut}) and (\ref{vbfcut}) with $\Delta\eta_{jj\rm min}=4$ 
 are imposed.
 For the massive-graviton productions, the additional $p_{T_j}$ cut
 (\ref{ptcut}) is also imposed.
 The distributions for each subprocess with the full diagrams (solid
 lines) and with the 
 only VBF diagrams (dashed lines) are shown.\label{fig:dphi}}}

To examine the validity of the above analytic parton-level expectations,
we plot in fig.~\ref{fig:dphi} the normalized azimuthal correlations
$\Delta\phi_{12}$ (mod $2\pi$) between the two tagging jets in the Higgs
+ 2-jet productions at the LHC, where the selection cuts (\ref{mincut})
and (\ref{vbfcut}) with $\Delta\eta_{jj\rm min}=4$ are applied. 
The distributions for each subprocess with the full diagrams and those
with the VBF diagrams only are shown by solid and dashed lines,
respectively. 
The VBF contributions can reproduce the distributions with the exact
matrix elements very well not only for the WBF processes but also for
the GF processes. As mentioned in the introduction, these azimuthal angle
correlations predicted in the leading order may survive even after
higher-order corrections are applied~\cite{DelDuca:2006hk,Andersen:2008gc}. 

It may be worth pointing out here that,
even though our definition of the azimuthal angles of the jets, which
are measured along the vector-boson colliding 
axis, is different from the usual definition along the beam axis in the
laboratory frame, the $\Delta\phi_{12}$ distributions
are almost same in the two frames due to the characteristic VBF kinematics.  
The VBF amplitudes are dominant in the collinear limit 
(\ref{cos1z1t}), where the vector-boson colliding axis 
({\it i.e.} the $z$-axis in fig.~\ref{fig:frame}) is identical with the beam
axis, and hence the $\Delta\phi_{12}$ distributions are not so much
distorted. In fact, our $\Delta\phi_{12}$ distributions for the Higgs
boson productions semi-quantitatively confirm those in the previous 
works~\cite{Plehn:2001nj,DelDuca:2001eu,Hankele:2006ja,Klamke:2007cu}.

Before turning to the spin-2 case, there are a
few remarks related to the previous studies on the azimuthal
correlations in the $Hjj$ events. 
(i)~The $XVV$ coupling form factors are factorized as in
eq.~(\ref{sigmahat}), and therefore, the $\Delta\phi_{12}$ distribution
is insensitive to their effects~\cite{Figy:2004pt}.
(ii)~As the ratio of the Higgs boson mass $M$ to the partonic
CM energy $\sqrt{\hat s}$ decreases, the interference
effect grows~\cite{Figy:2004pt}. In the collinear limit (\ref{cos1z1t}), 
we obtain 
\begin{align}
  \frac{M^2}{\hat s}
 =\frac{-4\cos\theta_1\cos\theta_2}{(1+\cos\theta_1)(1-\cos\theta_2)}
 =z_1z_2.
\end{align}
Therefore, as $M/\sqrt{\hat s}$ becomes smaller, {\it i.e.} 
$z_1z_2\to 0$, $\cos\theta_1$ and $\cos\theta_2$ approach zero, and the
ratio of $F_9^-$ to $F_1$ in eq.~(\ref{dphiGF}) grows;
see eqs.~(\ref{F1}) and (\ref{F9-}).
(iii)~Although we have considered the three types of tensor structures
separately for the Higgs coupling to vector bosons in this paper, it is
easy to extend our analyses to a mixed $CP$
scenario~\cite{Hankele:2006ja,Klamke:2007cu,Hankele:2006ma}; for
instance, the additional phases, which come from the $CP$-mixed $XVV$
coupling, can give rise to the {\it sine} terms in eq.~(\ref{dphidep0}),
and explain the shift of the dip positions in fig.~8 of
ref.~\cite{Klamke:2007cu}.

\subsubsection{Massive graviton productions}\label{sec:azcor_G}

Here, we discuss the case for the spin-2 particle productions, which is
more involved than the scalar case, because all the nine amplitudes
generically contribute to the total amplitude in eq.~(\ref{pamp}), which
can lead all the 25 azimuthal distributions in
eq.~(\ref{general_azdis}).  
Moreover, the graviton polarization along the momentum direction of the
decay products ($\lam'$) depends on the decay angle $\Theta$. 

In section~\ref{sec:VBFvsfull} we demonstrated that the QCD VBF
amplitudes can have significant contribution to the $G$ + 2-jet events
by imposing the VBF cuts and the $p_{T_j}$ slicing cut. 
In this case, the two off-shell VBF amplitudes, $\Mh_G{}_{+-}^{+2}$ and
$\Mh_G{}_{-+}^{-2}$, are dominant, as mentioned in
section~\ref{sec:vbfamp}. 
Therefore, only the $F_9^+\,(\propto\P^{+-}_{-+})$ term in
eq.~(\ref{general_azdis}) dominantly gives the nontrivial azimuthal
correlation, 
\begin{align}
 d\hat\sigma_{G}
 \sim \big[F_1^{}+2\,F_9^{+}\cos2\Phi_{12}\big]
\label{phidep2}
\end{align}
with $\Phi_{12}\equiv\phi_1+\phi_2$.
It should be emphasized here that $\Phi_{12}$ is not the azimuthal
separation $\Delta\phi_{12}\,(=\phi_1-\phi_2)$, but 
the sum of the azimuthal angles of
the two jets, $\phi_1$ and $\phi_2$.

From (\ref{phidep2}), one can immediately conclude that the
$\Delta\phi_{12}$ distributions for the
massive-graviton productions are flat.
In fig.~\ref{fig:dphi}, the $\Delta\phi_{12}$ correlations for the KK
graviton productions are also plotted with thick lines, where the
$p_{T_j}$ slicing cut~(\ref{ptslicingcut}) has been imposed as well as
the inclusive cuts~(\ref{mincut}) and the VBF cuts~(\ref{vbfcut}).%
\footnote{See also fig.~2 in ref.~\cite{Hagiwara:2008jb}, where $M_G=1$
TeV and the different selection cuts are applied.} 
The contributions from the $\lambda=\pm 1$ and 0 states, which can give
rise to the $\cos\Delta\phi_{12}$ and $\cos 2\Delta\phi_{12}$
dependence in eq.~(\ref{general_azdis}), are invisibly small; they are
smaller than the $\lambda=\pm 2$ by two and three orders of magnitude,
respectively. 
The flat $\Delta\phi_{12}$ distribution for the massive-graviton
productions is distinct from that for the SM Higgs boson productions,
which is expected to have a dip around $\Delta\phi_{12}=\pi/2$ due to
the GF contributions. 

Now let us see the explicit forms of the functions $F_1$ and $F_9^+$ in
eq.~(\ref{phidep2}). 
From eq.~(\ref{F1to9}) and by
using the explicit forms of the amplitudes in tables~\ref{Jq}(top), 
\ref{Jg}(top) and \ref{MG}, one finds
\begin{align}
 \hat F_1^G[qq/qg/gg]&=\hat F_1^H[qq/qg/gg]
  \times \frac{1}{2}\big\{\big(d^{\,2}_{+2,\lambda'}(\Theta)\big)^2
        +\big(d^{\,2}_{-2,\lambda'}(\Theta)\big)^2\big\}, 
\label{F1G}\\
 \hat F_9^+{}^G[qq/qg/gg]&=\hat F_9^-{}^H[qq/qg/gg]
  \times d^{\,2}_{+2,\lambda'}(\Theta)\,
         d^{\,2}_{-2,\lambda'}(\Theta),
\label{F9+G}  
\end{align}
where $\hat F_1^H$ and $\hat F_9^-{}^H$ are the same as in
eqs.~(\ref{F1}) and (\ref{F9-}), respectively, for the Higgs boson
productions, and the common overall factor is
\begin{align}
  F_{i}^{(\pm)}
 =\frac{2g_s^4g_{Ggg}^2M^4Q_1^2Q_2^2}{\cos^2\theta_1\cos^2\theta_2}\,
 \hat F_{i}^{(\pm)}{}^G.
\end{align}
Here, from table~\ref{MG}, we take $\Mh_G{}_{\pm\mp}^{\pm2}=-M^2$ in the
collinear limit.
By explicit forms of $J=2$ $d$ functions (see
table~\ref{dfunc}), the functions in eqs.~(\ref{F1G}) and (\ref{F9+G})
are 
\begin{align}
 \hat F_1^{G}[qq/qg/gg]=\hat F_1^H[qq/qg/gg] \times 
 \begin{cases}
  \frac{1}{16}(1+6\cos^2\Theta+\cos^4\Theta) &\text{for}\ \lam'=\pm2,\\
  \frac{1}{4}(1-\cos^4\Theta) &\text{for}\ \lam'=\pm1,\\
  \frac{3}{8}\sin^4\Theta &\text{for}\ \lam'=0,
 \end{cases}
\end{align}
and
\begin{align}
 \hat F_9^+{}^{G}[qq/qg/gg]=\hat F_9^-{}^H[qq/qg/gg] \times 
 \begin{cases}
  \frac{1}{16}\sin^4\Theta &\text{for}\ \lam'=\pm2,\\
  -\frac{1}{4}\sin^4\Theta &\text{for}\ \lam'=\pm1,\\
  \frac{3}{8}\sin^4\Theta &\text{for}\ \lam'=0.
 \end{cases}
\end{align}
The distributions strongly depend on the decay angle $\Theta$ and the
final polarization $\lambda'$.
The $\Theta$ dependence is the same for $\lam'=+2$ and $-2$,
and also for $\lam'=+1$ and $-1$.
Since $J=0$ resonances do not have such $\Theta$ dependence, the $J=2$
and $\lam'=0$ state can be distinguished from the $J=\lam'=0$
state in principle.
The coefficient function $F_9^+$ for all the final polarization states
is proportional 
to $\sin^4\Theta$, and hence
at $\Theta=0$ and $\pi$, where the decay axis 
 ($z'$-axis) is coincide with the initial polarization axis of the
 gravitons ($z$-axis), the azimuthal correlation is absent.
This is because each colliding vector-boson has the definite helicity.
Meanwhile, the correlation becomes larger with the larger decay
angle, and reaches the maximum at $\Theta=\pi/2$, where the $\lambda=+2$
and $-2$ states are mixed maximally.
It should be noted that the sign of the $F_{9}^+$ term
for $\lam'=\pm1$ is different for the other states, which gives rise to
the distinctive correlation. 

\FIGURE[t]{
 \epsfig{file=phi.eps,width=1\textwidth,clip}
 \caption{Azimuthal correlations $\Phi_{12}$ (mod $2\pi$) between the
 two tagging jets for the $qq$-scattering subprocess in $Gjj$
 productions at the LHC, where the same selection cuts in
 fig.~{\ref{fig:dphi}} are imposed. 
 The distributions for each final $G$ polarization state, $\lam'=\pm
 2$~(a), $\pm 1$~(b), and $0$~(c), are shown at the decay angle
 $\Theta=\pi/2$ (dashed), $\pi/3$ (dashed-dotted), and $\pi/6$ (dotted),
 where they are normalized by $\sigma(pp\to jjG)\,B(G\to a'_1a'_2)$. 
 The distributions after integrating out $\Theta$ are also shown by
 thick solid lines.
 \label{fig:phi}}}

After integrating out the decay angle $\Theta$, we obtain
\begin{align}
 \hat F_1^{G}[qq/qg/gg]&=\hat F_1^H[qq/qg/gg]
  \times \frac{2}{5}\{1,\,1,\,1\} &\text{for}\ \lam'=\{\pm2,\pm1,0\}, \\
 \hat F_9^+{}^{G}[qq/qg/gg]&=\hat F_9^-{}^H[qq/qg/gg]
  \times \frac{1}{15}\{1,\,-4,\,6\} &\text{for}\ \lam'=\{\pm2,\pm1,0\}.
\end{align}
The above results show that the magnitude of the azimuthal correlation
depends on the final polarization $\lambda'$.
The interference effect for the $\lambda'=0$ case is largest among the
possible five polarization states, 
although the decay branching ratio of KK gravitons into the $\lam'=0$
state, such as a pair of longitudinal weak-bosons and a $t\bar t$ pair with
same helicities, is less than 1\% for a whole mass range of gravitons. 
The correlation for $\lambda'=\pm1$, which is realized in the decays into a
fermion pair, is four times larger than that for $\lambda'=\pm2$, which
is projected out in the $G\to VV$ decays.  
It is worth noting that the above results depend only on 
$\lam'$, not on the decay mode. This universality of the
angular correlation can be an experimental signal of the $X$ spin
measurement. 

To examine the above analytic expectations, we demonstrate in 
fig.~\ref{fig:phi} the azimuthal correlations $\Phi_{12}$ (mod $2\pi$)
between the two tagging jets for the $qq$-collision subprocess in $Gjj$
events at the LHC, where the full diagrams are taken into account and
 the same selection cuts in fig.~{\ref{fig:dphi}} are imposed, 
{\it i.e.}, the selection cuts (\ref{mincut}), (\ref{vbfcut}) and
(\ref{ptcut}). 
 The distributions for each final polarization state, 
 $\lam'=\pm 2$, $\pm 1$, $0$, are shown at the decay angle $\Theta=\pi/2$,
 $\pi/3$, and $\pi/6$, where they are 
 normalized by $\sigma(pp\to jjG)\,B(G\to a'_1a'_2)$. 
 The distributions after integrating out $\Theta$ are also shown by
 thick solid lines, which is normalized to unity.
The above analytical results can describe the simulations very well.
For the $qg$- and $gg$-scattering cases, all the qualitative behaviors
are same, but the interference effects diminish as in the 
$\Delta\phi_{12}$ correlations for the Higgs boson case in fig.~\ref{fig:dphi}.
It should be stressed here that the observation of the $d$ function behavior,
or the $\Theta$ dependence, is a measurement of the $X$ spin, which can
be strengthened by the azimuthal correlation between the tagging jets.

\subsection{Correlations in the decay into a vector-boson pair}
\label{sec:az_dec}

As mentioned in section~\ref{sec:formalism}, the processes of a 
heavy-particle decay into a vector-boson pair which subsequently decay into 
$\ell\bar\ell$, $q\bar q$, or $gg$ are closely related to the VBF
production processes. Here we discuss the $X$ decay correlations between the
two decay planes of the vector bosons
by using the explicit helicity amplitudes, as in the production process.

To simplify the production part, we consider $s$-channel $X$
productions in $gg$ fusion or $q\bar q$ annihilation and its subsequent
decays into four-body final states, namely 
$n=0$ and $n'=4$ in (\ref{M2}), so that the initial polarization of $X$
along the $z$-axis can be fixed, $\lambda=\pm2$, $\pm1$, or $0$. 
In the following we take $\lambda=\bar\lambda$ and suppress the production
density matrix in eq.~(\ref{M2}).  

For the $X$ decay processes (\ref{x_jjjj}), the decay density matrix is given in 
eq.~(\ref{DtoDtensor}) in terms of the decay tensor
$\D^{\lam'_1\lam'_2}_{\bar\lam'_1\bar\lam'_2}$ of eq.~(\ref{dtensor}).
Similar to the azimuthal angle distributions for the production in 
eq.~(\ref{general_azdis}), those for the decays into a vector-boson pair
in eq.~(\ref{M2}) are generally expressed by~\cite{Hagiwara:1986vm} 
\begin{align}
  d^J_{\lam,\lam'}d^J_{\lam,\bar\lam'}
  \D^{\lam'_1\lam'_2}_{\bar\lam'_1\bar\lam'_2}
 = F'_1+\big\{2\,{\Re}e&\big[F'_2\cos\phi'_1+F'_3\cos\phi'_2
  +F'_4\cos2\phi'_1+F'_5\cos2\phi'_2 \nn\\
 &+F'_6{}^{\pm}\cos(\phi'_1\pm\phi'_2)
  +F'_7{}^{\pm}\cos(2\phi'_1\pm\phi'_2)+F'_8{}^{\pm}\cos(\phi'_1\pm2\phi'_2) \nn\\
 &+F'_9{}^{\pm}\cos2(\phi'_1\pm\phi'_2)\big]
  +({\Re}e\to{\Im}m,\,\cos\to\sin)\big\}.
\label{general_azdis_d}
\end{align}
Here, and in the following, summation over repeated indices 
$(\lam'_1,\lam'_2,\bar\lam'_1,\bar\lam'_2)=\pm,0$ 
is implied, and 
$\phi'_i$ is the azimuthal angle between the decay
plane of the vector boson and the $X$ production plane 
($gg/q\bar q\to X\to V'V'$) in the partonic CM frame; see also
fig.~\ref{fig:frame}. 

Because the azimuthal angle dependences of the quark and gluon current
amplitudes are the same for each helicity combination (see
tables~\ref{Jq}(bottom) and \ref{Jg}(bottom)), and because the phase of
the product of the two currents for $\sigma'_1=-\sigma'_3$ and 
$\sigma'_2=-\sigma'_4$ is (see eqs.~(\ref{J2qout_con}) and (\ref{J2gout_con}))
\begin{align}
 \Jh'_1{}^{\lam'_1}_{\sig'_1,\sig'_3=-\sig'_1}
 \Jh'_2{}^{\lam'_2}_{\sig'_2,\sig'_4=-\sig'_2}\propto
 e^{i(\lam'_1\phi'_1-\lam'_2\phi'_2)},
\label{J1J2phase_d}
\end{align}
the coefficients $F'{}_{\!\!\!1-9}^{(\pm)}$ for
$(f\bar f)(f\bar f)$, $(f\bar f)(gg)$ and $(gg)(gg)$ decays
are given in terms of two $d$ functions and the decay tensors
$\D^{\lam'_1\lam'_2}_{\bar\lam'_1\bar\lam'_2}$ as
\begin{align}
 F'_1&=
   d^J_{\lam,\lam'_1-\lam'_2}d^J_{\lam,\lam'_1-\lam'_2}
   \D_{\lam'_1\lam'_2}^{\lam'_1\lam'_2}, \nn\\
 F'_2&=
   d^J_{\lam,-\lam'_2}d^J_{\lam,1-\lam'_2}\D_{+\lam'_2}^{0\lam'_2}
  +d^J_{\lam,-1-\lam'_2}d^J_{\lam,-\lam'_2}\D_{0\lam'_2}^{-\lam'_2}, \nn\\
 F'_3&= 
   d^J_{\lam,\lam'_1-1}d^J_{\lam,\lam'_1}\D_{\lam'_10}^{\lam'_1+}
  +d^J_{\lam,\lam'_1}d^J_{\lam,\lam'_1+1}\D_{\lam'_1-}^{\lam'_10}, \nn\\
 F'_4&=
  d^J_{\lam,-1-\lam'_2}d^J_{\lam,1-\lam'_2}\D_{+\lam'_2}^{-\lam'_2}, \nn\\
 F'_5&=
  d^J_{\lam,\lam'_1-1 }d^J_{\lam,\lam'_1+1 }\D_{\lam'_1-}^{\lam'_1+}, \nn\\
 F'_6{}^{\pm}&= 
  d^J_{\lam,-1/1 }d^J_{\lam,1 }\D_{+0}^{0\pm}
            +
  d^J_{\lam,0 }d^J_{\lam,2/0 }\D_{+\mp}^{00}
            +
  d^J_{\lam,-2/0 }d^J_{\lam,0 }\D_{00}^{-\pm}
            +
  d^J_{\lam,-1 }d^J_{\lam,1/-1 }\D_{0\mp}^{-0}, \nn\\
 F'_7{}^{\pm}&= 
  d^J_{\lam,-2/0 }d^J_{\lam,1 }\D_{+0}^{-\pm}
            +
  d^J_{\lam,-1 }d^J_{\lam,2/0 }\D_{+\mp}^{-0}, \nn\\
 F'_8{}^{\pm}&= 
  d^J_{\lam,-1/1 }d^J_{\lam,2/0 }\D_{+\mp}^{0\pm}
            +
  d^J_{\lam,-2/0 }d^J_{\lam,1/-1 }\D_{0\mp}^{-\pm}, \nn\\
 F'_9{}^{\pm}&=
  d^J_{\lam,-2/0 }d^J_{\lam,2/0 }\D_{+\mp}^{-\pm}. 
\label{F'1to9}
\end{align}
We note that this relations can be also applied to 
the processes with the same-helicity currents, namely 
the cases for $\sigma'_1=\sigma'_3$ and/or $\sigma'_2=\sigma'_4$, which
appear only in the $g^*\to gg$ currents; the same arguments in
the production part can be applied (see below eq.~(\ref{F1to9})).
We should note that, similar to eq.~(\ref{F1to9}) for the VBF production
processes, the $\Theta$-dependent azimuthal angle correlations are
manifestly expressed 
by quantum interference among different helicity states of the
intermediate vector-bosons in eq.~(\ref{F'1to9}). 

\subsubsection{Higgs boson decays}

The decay amplitudes~(\ref{damp}) for scalar particles are the
coherent sum of the three amplitudes in which the decaying vector-bosons
have the same helicities ($\lambda'_1=\lambda'_2$), and there is no
$\Theta$ dependence, {\it i.e.} $d^{J=0}_{\lam,\lam'}(\Theta)=1$ in
eq.~(\ref{general_azdis_d}) (and in eq.~(\ref{F'1to9})). 
Therefore, similar to the production case, the azimuthal angle
correlation~(\ref{general_azdis_d}) for the $J=0$ particle decays is
reduced to  
\begin{align}
  \D^{\lam'_1\lam'_2}_{\bar\lam'_1\bar\lam'_2}
 = F'_1+\big\{2\,{\Re}e\big[F'_6{}^{-}\cos\Delta\phi'_{12}
  +F'_9{}^{-}\cos2\Delta\phi'_{12}\big]
  +({\Re}e\to{\Im}m,\,\cos\to\sin)\big\},
\label{dphidep0_d}
\end{align}
where $\Delta\phi'_{12}\equiv\phi'_1-\phi'_2$ is the
angle between the two decay planes of the vector bosons.

The decay angular correlations have been extensively studied for
the Higgs boson decays into the weak bosons, $H\to Z^{(*)}Z$,
$W^{(*)}W$~\cite{Dell'Aquila:1985ve,Skjold:1993jd,Hohlfeld:2001,Choi:2002jk,Buszello:2002uu}.
It may be worth pointing that the $F'_6{}^-$ and $F'_9{}^-$ terms in
eq.~(\ref{dphidep0_d}) can be observable for the Higgs boson mass
$M_H<400$~GeV since the decays into a pair of 
transverse weak-bosons are not negligible compared to the decays into
longitudinal ones, as shown in~(\ref{HVVdecay}).
$F'_6{}^-$ is very small for the $H\to ZZ$ decay, while it is large for
the $H\to WW$ decay due to the $P$-odd
nature~\cite{Skjold:1993jd,Djouadi:2005gi}. 
One can see detailed simulations for the LHC in 
refs.~\cite{Hohlfeld:2001,Choi:2002jk,Buszello:2002uu}. 
It must be recalled here that there is no azimuthal
correlation between the two jets in the WBF Higgs productions as
discussed in eq.~(\ref{dphiWBF}).

Here we discuss the correlations for the Higgs boson decays into a
virtual-gluon pair, 
$H/A\to g^*g^*\to(q\bar q)(q\bar q)$, $(q\bar q)(gg)$,
or $(gg)(gg)$, in some detail. In the on-shell limit, or 
$Q'_{1,2}\to 0$, where the above decay processes via a virtual-gluon pair
are dominant, 
only the transverse amplitudes ($\Mh'_{H/A}{}_{\pm\pm}$)
contribute to the decay amplitudes both
for the $CP$-even/odd Higgs boson decays.
Moreover, due to the $P$-odd nature, the $F'_6{}^-$ term vanishes for the
QCD processes as in the VBF processes.
Therefore, only the $F'_9{}^-$ term in eq.~(\ref{dphidep0_d}) is
relevant in this case.

From the relation in (\ref{F'1to9}) and by using the explicit forms of
the amplitudes in tables~\ref{Jq}(bottom), \ref{Jg}(bottom) and
\ref{Mh}, the coefficients $F'_1$ and $F'_9{}^-$ in
eq.~(\ref{dphidep0_d}) for each decay process, $(q\bar q)(q\bar q)$,
$(q\bar q)(gg)$ and $(gg)(gg)$, are expressed (except for the color factors) as 
\begin{subequations}
\begin{align}
 \hat F'_1[(q\bar q)(q\bar q)]&=(1+\cos^2\theta'_1)(1+\cos^2\theta'_2), \\
 \hat F'_1[(q\bar q)(gg)]&=\dfrac{1}{\sin^2\theta'_2}
                   (1+\cos^2\theta'_1)(3+\cos^2\theta'_2)^2 
                  \quad {\rm or}\quad (1\leftrightarrow 2), \\
 \hat F'_1[(gg)(gg)]&=\dfrac{1}{\sin^2\theta'_1\sin^2\theta'_2}
                   (3+\cos^2\theta'_1)^2(3+\cos^2\theta'_2)^2,
\end{align}
\label{F1'}%
\end{subequations}
and
\begin{subequations}
\begin{align}
  \hat F'_9{}^-[(q\bar q)(q\bar q)/(gg)(gg)]
 &=\pm\frac{1}{2}(1-\cos^2\theta'_1)(1-\cos^2\theta'_2), \\
  \hat F'_9{}^-[(q\bar q)(gg)]
 &=\mp\frac{1}{2}(1-\cos^2\theta'_1)(1-\cos^2\theta'_2),
\label{F9'-}
\end{align}
\end{subequations}
where the upper (lower) sign corresponds to the $CP$-even (-odd) Higgs
boson case. 
Here we take the $Q'_1,Q'_2\ll M$ limit for the $X\to VV$ amplitudes,
where the only surviving amplitudes are 
$\big|\Mh'_{H/A}{}_{\pm\pm}\big|=\frac{1}{2}M^2$, with the
common overall factor 
\begin{align}
  F'_{i}{}^{(\pm)}
 =\frac{1}{2}g_{H/Agg}^2g_s^4M^4Q'_1{}^2Q'_2{}^2\,
  \hat F'_{i}{}^{(\pm)}.
\end{align}
The constant term $F'_1$ has the different contributions from the
$q\bar q$ decay and the $gg$ decay and can be larger as the process
involves the $g^*\to gg$ splitting. 
As in the production processes with the helicity-flip currents,
the same-helicity currents ($\sig'_1=\sig'_3$ and/or
$\sig'_2=\sig'_4$), which appear only in the gluon currents,
give rise only the constant piece, $F'_1$. 
The interference term $F'_9{}^-$, which receives the
contribution only from the opposite-helicity currents
($\sig'_1=-\sig'_3$ and $\sig'_2=-\sig'_4$), is same except for sign in
all the decay processes. 
This is because the relative sign between the different helicity states
for the outgoing fermion and
gluon current amplitudes is different; see tables~\ref{Jq}(bottom) and 
\ref{Jg}(bottom), and because the relative phases between
$\Jh'_1{}^{\lambda'}_{\sig'_1,-\sig'_1}$ and $\Jh'_2{}^{\lambda'}_{\sig'_2,-\sig'_2}$ for the quark currents 
(\ref{J2qout_con}) are different from those for the gluon currents
(\ref{J2gout_con}). 

Aside from the coupling constant, the double $(q\bar q)$ decays via a pair of
gluons are the same as the decays via a virtual-photon pair, 
$H/A\to \gamma^*\gamma^*\to(\ell\bar\ell/q\bar q)(\ell\bar\ell/q\bar q)$, 
and we expect the strong correlations as in the GF $qq\to qqH/A$ processes.
On the other hand, the decay correlations for $(q\bar q)(gg)$ and
$(gg)(gg)$ are much smaller than those for $qg\to qgH/A$ and 
$gg\to ggH/A$ since
the reduction factor from the gluon currents for the decays, 
$(3+\cos^2\theta'_i)^2$ in~(\ref{F1'}), can be much 
larger than that for the productions, $(1+3\cos^2\theta_i)^2$ in~(\ref{F1}).
This is well known as the knowledge of the QCD parton branching; a
quark-antiquark pair from the gluon, $g^*\to q\bar q$, has the strong
correlation between the gluon polarization and the decay plane, while
for $g^*\to gg$ the correlation is weak~\cite{Ellis:1991qj}. 
It may be a challenging task to observe the azimuthal correlations between the QCD
jets in the Higgs boson decays at the LHC. 
The quantitative study will be reported elsewhere.

\subsubsection{Massive graviton decays}

For the spin-2 particle decays, unlike the scalar particle decays, the nine
amplitudes should be coherently summed  in eq.~(\ref{damp}) and may give
rise to all the 25 azimuthal distributions in
eq.~(\ref{general_azdis_d}).   

In this article, we study the process of the graviton decays into a
weak-boson pair at the LHC,
\begin{align}
 pp\to G\to WW/ZZ\to (f_1\bar f_3)(f_2\bar f_4).
\end{align}
This includes final four-charged-lepton signals,  
which may give a clean signal and allow a complete
kinematical reconstruction at the LHC.
The $s$-channel $G$ production has two possible sources, $gg$ fusion and
$q\bar q$ annihilation, even though the $gg$ contribution dominates the cross
section for the graviton mass up to 3.4 TeV~\cite{Allanach:2002gn}.   
In the parton CM frame, the polarization of the produced gravitons is
fixed along the beam axis ($z$-axis) at $\lambda=+2$ or $-2$ in 
gluon fusion, while at $\lambda=+1$ or $-1$ in $q\bar q$
annihilation. The two different production modes lead
totally different angular $\Theta$ distributions, as we will see below.  

As mentioned in section~\ref{sec:vbfamp}, in the heavy graviton mass limit
($\beta'=\sqrt{1-4m_V^2/M^2}\to1$), only three $G\to VV$
amplitudes survive for the
decays into on-shell weak-bosons; see table~\ref{MGVV}.
In this limit, therefore, the azimuthal distributions in
eq.~(\ref{general_azdis_d}) can be reduced to  
\begin{align}
  d^2_{\lam,\lam'}d^2_{\lam,\bar\lam'}  
  \D^{\lam'_1\lam'_2}_{\bar\lam'_1\bar\lam'_2}
 = F'_1+2\,F'_6{}^{+}\cos\Phi'_{12}
  +2\,F'_9{}^{+}\cos2\Phi'_{12}
\label{dphidep2_d}
\end{align}
with $\Phi'_{12}\equiv\phi'_1+\phi'_2$. 
It should be stressed here that
$\Phi'_{12}$ is not the angle between two decay planes 
$\Delta\phi'_{12}\,(=\phi'_1-\phi'_2)$, but the sum of  $\phi'_1$ and $\phi'_2$, 
which are measured separately from the graviton production 
plane.

From eq.~(\ref{F'1to9}) and by using the explicit forms of the helicity
amplitudes in tables~\ref{Jq}(bottom) and \ref{MGVV} and $d$~functions in
table~\ref{dfunc}, the complete angular distributions of 
eq.~(\ref{dphidep2_d}) are presented for $\lambda=\pm2$ and $\pm1$
in appendix~\ref{sec:gravitondecay}.  
After integrating out $\theta'_1$ and $\theta'_2$ in (\ref{Gdecay2}) and
(\ref{Gdecay1}), the $\Theta$-dependent coefficients in
eq.~(\ref{dphidep2_d}) are 
\begin{subequations}
\begin{align}
 \hat F'_1 &= 
  3+10\cos^2\Theta+3\cos^4\Theta, \\
 \hat F'_6{}^+ &=
  \kappa_1\kappa_2\,\frac{9\pi^2}{64}(1-\cos^4\Theta), \\ 
 \hat F'_9{}^+ &=
  \frac{1}{4}\sin^4\Theta,
\end{align}
\label{F'i_l2}%
\end{subequations}
for $\lambda=\pm2$ (the $gg$ initial state), and 
\begin{subequations}
\begin{align}
 \hat F'_1 &= 
  8+4\cos^2\Theta-12\cos^4\Theta, \\
 \hat F'_6{}^+ &=
  -\kappa_1\kappa_2\,\frac{9\pi^2}{16}\sin^2\Theta\cos^2\Theta, \\
 \hat F'_9{}^+ &=
  -\sin^4\Theta,
\end{align}
\label{F'i_l1}%
\end{subequations}
for $\lambda=\pm1$ (the $q\bar q$ initial state), where the common over
all factor is 
\begin{align}
  F'_{i}{}^{(\pm)}
 =\frac{1}{18}g_{GVV}^2g_{V_1f_1f_3}^2g_{V_2f_2f_4}^2
  M^4{Q'_1}^2{Q'_2}^2\,
 \hat F'_{i}{}^{(\pm)}
\end{align}
with
$g_{V_if_if_{i+2}}^2= \big(g_-^{V_if_if_{i+2}}\big)^2
                     +\big(g_+^{V_if_if_{i+2}}\big)^2$, 
and the combination of the $Vff$ couplings $\kappa_i$ is
\begin{align}
 \kappa_i =
  \big[\big(g_-^{V_if_if_{i+2}}\big)^2-\big(g_+^{V_if_if_{i+2}}\big)^2
  \big]\big/g_{V_if_if_{i+2}}^2. 
\label{eta}
\end{align}
The $Vff$ couplings are given in (\ref{g_Vff}).
The $\Theta$ distributions for the $\Phi'_{12}$-independent term, $F'_1$,
are totally different between the two initial states due to the angular
momentum conservation; a pair of vector bosons from the $\lambda=\pm2$
gravitons tend to decay to
the forward and backward directions, while the $\lambda=\pm1$ gravitons
are not allowed to decay into a vector-boson pair at $\Theta=0$ and
$\pi$ (in the $\beta'=1$ limit), where the momentum direction of the decaying 
vector-bosons ($z'$-axis) is coincide with the polarization axis of the
gravitons ($z$-axis). 
On the other hand, all the azimuthal correlations, $F'_6{}^+$ and
$F'_9{}^+$, disappear at $\Theta=0$
and $\pi$ since each decaying vector-boson has the definite helicity. 
The coefficient $F'_6{}^+$ takes the maximum at $\Theta=\pi/2$
($\pi/4$) for $\lambda=\pm2$ ($\pm1$), while the $F'_9{}^+$ becomes 
larger with the larger decay angle and reaches
the maximum at $\Theta=\pi/2$ both for $\lambda=\pm2$ and $\pm1$. 
The sign difference of the $\Phi'_{12}$-dependent terms between 
$\lambda=\pm 2$ and $\pm 1$ gives rise to the distinctive correlations.

Let us estimate the asymmetries $A_i\equiv 2 F'_i{}^+/F'_1$ ($i=6,9$)
from (\ref{F'i_l2}) and (\ref{F'i_l1}) for the $G\to ZZ\to 4\ell$
process. After the integration for $\Theta$, one finds 
\begin{align}
 A_6 &= \kappa_1\kappa_2\, \frac{27\pi^2}{832}\sim 0.007, &
 A_9 &= \frac{1}{26}\sim 0.038 
 &{\rm for}\ \lambda=\pm 2, 
\label{Afor2}\\
 A_6 &= -\kappa_1\kappa_2\, \frac{9\pi^2}{416}\sim -0.005, &
 A_9 &= -\frac{2}{13}\sim -0.154 
 &{\rm for}\ \lambda=\pm 1.
\label{Afor1}
\end{align}
The asymmetry $A_6$ of the $\cos\Phi'_{12}$ term is tiny both
for $\lam=\pm2$ and $\pm1$, less
than 1\%, due to the smallness of the parity violation for the $Z$
decays, similar to the $H\to ZZ\to 4\ell$ process. On the other hand, 
$A_9$ of the $\cos2\Phi'_{12}$ term reaches around 4\% for
$\lambda=\pm 2$ and 15\% for $\lambda=\pm 1$.

We simulate the $G\to ZZ\to 4\ell$ 
process to examine our analytic expectations, 
where we take $M=600$ GeV, $m_Z=91.2$ GeV and $\Gamma_Z=2.5$ GeV, that is, all
the nine amplitudes in (\ref{damp}) are taken into account without any
approximation. 
Figure~\ref{fig:phi_d} shows the 
normalized azimuthal correlations $\Phi'_{12}$ (mod $2\pi$) between
the decay planes of the vector bosons for $\lambda=\pm 2$ (a) and $\pm 1$ (b). 
The decay angle 
$\Theta$ is fixed at $\pi/2$, $\pi/3$, and $\pi/6$, shown by dashed, 
dashed-dotted, and dotted lines, respectively. 
The distributions
 after integrating out $\Theta$ are also shown by thick solid lines,
 which are normalized to unity. 
Our analytic approximation can explain the results not only
qualitatively but also quantitatively.
It should be stressed again that the observation of the $\Theta$
dependence of the azimuthal correlations can be a measurement of the $X$ spin.

\FIGURE[t]{
 \epsfig{file=phi_d.eps,width=0.7\textwidth,clip}
 \caption{Normalized azimuthal correlations $\Phi'_{12}$ (mod $2\pi$) between
 the decay planes of the vector bosons in the $G\to ZZ\to 4\ell$ decays
 for the initial helicity
   $\lambda=\pm 2$ (a) and $\pm 1$ (b), where the
 decay angle $\Theta$ is fixed at $\pi/2$ (dashed), $\pi/3$ (dashed-dotted), and 
 $\pi/6$ (dotted).
The distributions
 after integrating out $\Theta$ are also shown by thick solid lines.  
 \label{fig:phi_d}}}

At the LHC, as mentioned before, $gg$ fusion is the main 
production process of KK gravitons, that is, the $\lambda=\pm 2$
states are mainly produced. 
Therefore, the azimuthal distribution is expected to be suppressed around 
$\Phi'_{12}=\pi/2$.   
However, for the higher graviton mass region, the contribution from 
$q\bar q$ annihilation, which enhance the events around $\Phi'_{12}=\pi/2$, 
becomes larger and gives rise to cancel the correlation each
other.
At the graviton mass $M=3.4$ TeV the production rate of the two contributions is
comparable~\cite{Allanach:2002gn}, and hence the azimuthal distributions
are enhanced around $\Phi'_{12}=\pi/2$ since the correlation in 
$q\bar q$ annihilation is stronger than that in gluon fusion; see
(\ref{Afor2}) and (\ref{Afor1}).

In the future $e^+e^-$ (or photon) linear collider, on the other hand,
only the $\lambda=\pm 1$ (or $\lambda=\pm 2$) state can be produced, and 
more precise studied to determine the spin-2 nature may 
be done in the clean environment.

\section{Summary}\label{sec:summary}

We have studied angular correlations of the two accompanying jets in
Higgs boson and massive-graviton productions
at hadron colliders, which include WBF and GF processes.
We have also considered their decays into a vector-boson pair which
subsequently decay into $\ell\bar\ell$, $q\bar q$ or $gg$. 

The amplitudes for the VBF subprocesses are given by the product of the
two incoming current amplitudes and the off-shell $VV\to X$ amplitude
summed over 
the polarization of the $t$-channel intermediate vector-bosons (see
eq.~(\ref{pamp})), while the amplitudes for the $X$ decays into four
final states via a vector-boson pair are expressed as the product of the
$X\to VV$ amplitude and the two outgoing current amplitudes summed over
the polarization of the $s$-channel intermediate vector-bosons (see
eq.~(\ref{damp})). Using the kinematical variables in fig.~\ref{fig:frame},   
we presented all the helicity amplitudes explicitly; tables~\ref{Jq} and
\ref{Jg} for the quark and gluon currents, and tables~\ref{Mh} and
\ref{MG} for the $VV\to H/A$ and $VV\to G$ processes, respectively. 
We also showed  that our off-shell vector-boson current amplitudes reduce
to the standard quark and gluon splitting amplitudes with appropriate
gluon-polarization phases in the collinear limit (eqs.~(\ref{cos1z1t})
and (\ref{cos1z1s})). 

To validate our analyses, we demonstrated that the VBF amplitudes
dominate the exact matrix elements not only for the WBF processes but
also for all the GF processes when typical selection cuts to enhance
the VBF events are applied, such as a large rapidity separation between
two jets in (\ref{vbfcut}).
Furthermore, we found that the ${p_T}_j$ slicing cut
(\ref{ptslicingcut}) is effective to suppress the non-VBF diagrams
especially for the graviton productions. 

By using the density matrix formalism in eq.~(\ref{M2}) and our
analytical amplitudes, we showed that  
nontrivial azimuthal angle correlations of the jets in the
production and in the decay of massive spin-0 and -2 bosons are
manifestly expressed in terms of the quantum interference among
different helicity states of the intermediate vector-bosons; see 
(\ref{F1to9}) for the production and (\ref{F'1to9}) for the decay.

For the productions and the decays of Higgs bosons, our analytical
arguments can describe the previous studies on the angular correlations;
for instance, the WBF gives the flat azimuthal distribution
(\ref{dphiWBF}),  while the GF
produces the $\cos2(\phi_1-\phi_2)$ distribution (\ref{dphiGF}).
We also explicitly showed that the gluon currents, especially in the
decay processes, reduce the azimuthal correlations.
For the massive-graviton case, 
we found the $\Theta$-dependent $\cos2(\phi_1+\phi_2)$ correlations for both
the production processes (\ref{phidep2}) and the decay processes
(\ref{dphidep2_d}), 
which are proportional to $\sin^4\Theta$. The correlations also depend
on the final graviton polarization in the productions and on the initial
graviton polarization in the decays.
Those correlations reflect the spin and the $CP$ nature of the Higgs
bosons and the massive gravitons, and may have a great potential to be
observed at the LHC.

\acknowledgments
We wish to thank Y.~Matsumoto and D.~Nomura for discussions in the  
early stage of the investigation. 
Q.L.\ and K.M.\ would like to thank the KEK theory group for the warm
 hospitality, and also the IPMU (Institute for Physics and Mathematics
 of the Universe) for organizing LHC focus week meetings in December
 2007 and March 2009 where we enjoyed stimulating discussions.
K.H.\ and K.M.\ would like to thank the Aspen Center for Physics and program:
``LHC: Beyond the Standard Model Signals in a QCD Environment'' 
where a part of this work was done. 
K.M.\ also thanks O.~Nachtmann, T.~Plehn and S.~Schumann for valuable
comments. 
This work is supported in part by the Core University Program of JSPS,
and in part by the Grant-in-Aid for Scientific Research (No.~20340064) from
MEXT, Japan.

The diagrams in this paper were drawn using 
{\tt JaxoDraw}~\cite{Binosi:2003yf}.

\appendix
\section{Wavefunction and vertices for a spin-2 particle}
\label{sec:spin2} 

The polarization tensor for a spin-2 particle in table~\ref{VVX},
$\eps^{\mu\nu}(p,\lam)$, is decomposed into polarization vectors for a
spin-1 particle as  
\begin{align}
 \eps^{\mu\nu}(p,\pm 2) &= \eps^{\mu}(p,\pm)\,\eps^{\nu}(p,\pm), \nn\\
 \eps^{\mu\nu}(p,\pm 1) &= \frac{1}{\sqrt{2}}
               \big[ \eps^{\mu}(p,\pm)\,\eps^{\nu}(p,0)
                    +\eps^{\mu}(p,0)\,\eps^{\nu}(p,\pm)\big], \nn\\
 \eps^{\mu\nu}(p, 0) &= \frac{1}{\sqrt{6}}
               \big[  \eps^{\mu}(p,+)\,\eps^{\nu}(p,-)
                    + \eps^{\mu}(p,-)\,\eps^{\nu}(p,+)
                    +2\,\eps^{\mu}(p,0)\,\eps^{\nu}(p,0)\big].
\label{spin2wavefunc}
\end{align} 
See more details in ref.~\cite{Hagiwara:2008jb}.

The vertices for a massive graviton with two vector-bosons in
table~\ref{VVX} are given by~\cite{Giudice:1998ck}
\begin{align}
 \hat\Gamma^{\mu\nu,\alpha\beta}_{GVV}(q_1,q_2) 
 = (m_V^2+q_1\cdot q_2)\,C^{\mu\nu,\alpha\beta}
  +D^{\mu\nu,\alpha\beta}(q_1,q_2)
  +\xi^{-1}E^{\mu\nu,\alpha\beta}(q_1,q_2),
\end{align}
where $q_{1,2}$ and $m_V$ are the momenta and mass of the vector bosons,
and 
\begin{align}
 C^{\mu\nu,\alpha\beta} &= g^{\mu\alpha}g^{\nu\beta}+g^{\mu\beta}g^{\nu\alpha}
                         -g^{\mu\nu}g^{\alpha\beta}, \\
 D^{\mu\nu,\alpha\beta}(q_1,q_2) &= g^{\mu\nu}q_1^{\beta}q_2^{\alpha}
  -\big[ g^{\mu\beta}q_1^{\nu}q_2^{\alpha}+g^{\mu\alpha}q_1^{\beta}q_2^{\nu}
        -g^{\alpha\beta}q_1^{\mu}q_2^{\nu}+(\mu\leftrightarrow\nu)\big], \\ 
 E^{\mu\nu,\alpha\beta}(q_1,q_2) &= g^{\mu\nu}
   (q_1^{\alpha}q_1^{\beta}+q_2^{\alpha}q_2^{\beta}+q_1^{\alpha}q_2^{\beta})
  -\big[ g^{\nu\beta}q_1^{\mu}q_1^{\alpha}+g^{\nu\alpha}q_2^{\mu}q_2^{\beta}
        +(\mu\leftrightarrow\nu)\big].
\end{align}
The $\xi$ term is the gauge-fixing term, which vanishes for massive
vector-bosons in the unitary gauge ($\xi\to\infty$).
For massless vector-bosons we take $\xi=1$ in the Feynman gauge.

\section{Relation between wavefunctions and $d$~functions}
\label{sec:dfunc}

In this appendix, we demonstrate that two-to-two processes via
$s$-channel spin-2 resonances can be factorized into 
the production part and its decay part, by using explicit spin-2
wavefunctions.  
In the resonance ($X$) rest frame, $p^{\mu}=(m,\vec 0)$,
we consider the processes 
\begin{align}
 a_1(k_1,\sig_1)+a_2(k_2,\sig_2) \to X(p) \to 
 a'_1(k'_1,\sig'_1)+a'_2(k'_2,\sig'_2),
\end{align}
where the $a_1$ momentum ($k_1$) is taken along the positive $z$-axis
and the $a'_1$ momentum ($k'_1$) is given by the scattering angle
$\theta$.

Before we consider the spin-2 resonance case, let us start with the
well-known spin-1 case. 
The numerator of the propagator for spin-1 particles, 
\begin{align}
 P^{\mu\nu}(p) =g^{\mu\nu}-\frac{p^{\mu}p^{\nu}}{p^2}
 =-\sum_{\lam=\pm1,0}\eps^{\mu}(p,\lam)^*\,\eps^{\nu}(p,\lam),
\end{align}
is a projector on the on-shell particle, $p^2=m^2$.
Since the projector is an operator which satisfies $P^2=P$, namely
\begin{align}
 P^{\mu\nu}(p)&=P^{\mu\rho}(p)P_{\rho}^{\ \nu}(p) \nn\\ &=
  \sum_{\lam}\eps^{\mu}(p,\lam)^*\,\eps^{\rho}(p,\lam)
  \sum_{\lam'}\eps_{\rho}(p,\lam')^*\,\eps^{\nu}(p,\lam').
\label{projector1}
\end{align}
Here, $\lam$ is the helicity along the incoming $a_1$ momentum and
$\lam'$ is the helicity along the outgoing $a'_1$ momentum, that is,
$\lam=\sig_1-\sig_2$ and $\lam'=\sig'_1-\sig'_2$.
On the mass-shell,
the propagator factor~(\ref{projector1}) can be factorized into the
production part and the decay part; the wavefunction 
$\eps^{\mu}(p,\lam)^*$ is used to calculate the production amplitudes,
and the wavefunction $\eps^{\nu}(p,\lam')$ is used to calculate the
decay amplitudes. There is a connecting factor,
$\eps^{\rho}(p,\lam)\,\eps_{\rho}(p,\lam')^*$,
which should be a scalar function depending on $\lam$, $\lam'$, and
the orientation angle $\theta$ between the two quantization axes.
Using the explicit forms of the spin-1 polarization vectors
(with the {\tt HELAS} convention~\cite{Murayama:1992gi}):
\begin{align}
 \eps^{\mu}(p,\lam=\pm)&=\frac{1}{\sqrt{2}}(0,\,\mp1,\,-i,\,0), \nn\\
 \eps^{\mu}(p,\lam=0)  &=(0,\,0,\,0,\,1), 
\label{spin1wavefunc_i}
\end{align}
and
\begin{align}
 \eps^{\mu}(p,\lam'=\pm)&=\frac{1}{\sqrt{2}}
  (0,\,\mp\cos\theta,\,-i,\,\pm\sin\theta), \nn\\
 \eps^{\mu}(p,\lam'=0)&=(0,\,\sin\theta,\,0,\,\cos\theta),
\label{spin1wavefunc_f}
\end{align}
the connecting factor can be expressed as the $J=1$ $d$~function,
\begin{align}
  \eps^{\rho}(p,\lam)\,\eps_{\rho}(p,\lam')^*
 =-\begin{bmatrix}
   \frac{1}{2}(1+\cos\theta) & -\frac{1}{\sqrt{2}}\sin\theta & 
   \frac{1}{2}(1-\cos\theta) \\ 
   \frac{1}{\sqrt{2}}\sin\theta & \cos\theta & 
   -\frac{1}{\sqrt{2}}\sin\theta \\
   \frac{1}{2}(1-\cos\theta) & \frac{1}{\sqrt{2}}\sin\theta & 
   \frac{1}{2}(1+\cos\theta)
   \end{bmatrix} 
 =-d^1_{\lam,\lam'}(\theta).
\label{d1}
\end{align}
The $d$~function dictates the overlap of the angular momentum states
between the initial $a_1a_2$ and the final $a'_1a'_2$ state. 
Finally, the propagator factor (\ref{projector1}) can be rewritten as 
\begin{align}
 P^{\mu\nu}(p)
 =-\sum_{\lam,\lam'}\eps^{\mu}(p,\lam)^*\,
  d^1_{\lam,\lam'}(\theta)\,
  \eps^{\nu}(p,\lam').
\end{align}
Note that the sign comes from the orthogonal relation,
$\eps^{\mu}(p,\lam^{(\prime)})\,\eps_{\mu}(p,\lam^{(\prime)})^*=-1$.

Now, let us move to the spin-2 case.
Similarly, the propagator factor for spin-2 particles is a projector on
the mass-shell ($p^2=m^2$):
\begin{align}
 P^{\mu\nu\alpha\beta}(p)
  &= \frac{1}{2}
     ( g^{\mu\alpha}g^{\nu\beta}
      +g^{\mu\beta}g^{\nu\alpha}
      -g^{\mu\nu}g^{\alpha\beta} ) \nn\\
  &\quad-\frac{1}{2p^2}
     ( g^{\mu\alpha}{p^\nu p^\beta}+g^{\nu\beta}{p^\mu p^\alpha}
      +g^{\mu\beta}{p^\nu p^\alpha}+g^{\nu\alpha}{p^\mu p^\beta})
  \nn\\
  &\quad+\frac{1}{6}
   \Big(g^{\mu\nu}+\frac{2}{p^2}p^\mu p^\nu\Big)
   \Big(g^{\alpha\beta}+\frac{2}{p^2}p^\alpha p^\beta\Big) \nn\\
 &=\sum_{\lambda=\pm2,\pm1,0} \epsilon^{\mu\nu}(p,\lambda)^*
  \epsilon^{\alpha\beta}(p,\lambda),
\end{align}
where $\epsilon^{\mu\nu}(p,\lambda)$ is a spin-2 wavefunction. 
On the mass-shell, the summation over the helicity can be duplicated,
\begin{align}
 P^{\mu\nu\alpha\beta}(p) &= 
 P^{\mu\nu\rho\sig}(p)P_{\rho\sig}^{\ \ \alpha\beta}(p) \nn\\ &=
  \sum_{\lam}\eps^{\mu\nu}(p,\lam)^*\,\eps^{\rho\sig}(p,\lam)
  \sum_{\lam'}\eps_{\rho\sig}(p,\lam')^*\,\eps^{\alpha\beta}(p,\lam').
\label{projector2}
\end{align}
By using the explicit forms of the spin-2 polarization tensors in
eq.~(\ref{spin2wavefunc}) and the polarization vectors in 
eqs.~(\ref{spin1wavefunc_i}) and (\ref{spin1wavefunc_f}),
one finds that the overlap factor is the $J=2$ $d$~function (see
table~\ref{dfunc}) as
\begin{align}
  \eps^{\rho\sigma}(p,\lam)\,\eps_{\rho\sigma}(p,\lam')^* 
 =d^2_{\lam,\lam'}(\theta).
\end{align}
Therefore, the propagator factor for spin-2 particles (\ref{projector2})
can be factorized into the production and the decay part as
\begin{align}
 P^{\mu\nu\alpha\beta}(p)
 =\sum_{\lam,\lam'}\eps^{\mu\nu}(p,\lam)^*\,
  d^2_{\lam,\lam'}(\theta)\,
  \eps^{\alpha\beta}(p,\lam').
\end{align}

\TABULAR[t]{|cc|cc|}{\hline \\[-4mm]
 $d^{\,2}_{\pm 2,2}=d^{\,2}_{\mp 2,-2}$ &
  $\frac{1}{4}(1\pm\cos\theta)^2$ & & \\
 $d^{\,2}_{\pm 2,1}=-d^{\,2}_{\mp 2,-1}$ &
  $\mp\frac{1}{2}(1\pm\cos\theta)\sin\theta$ &
 $d^{\,2}_{\pm 1,1}=d^{\,2}_{\mp 1,-1}$ &
  $\frac{1}{2}(1\pm\cos\theta)(2\cos\theta\mp 1)$ \\
 $d^{\,2}_{\pm 2,0}$ &
  $\frac{\sqrt{6}}{4}\sin^2\theta$ &
 $d^{\,2}_{\pm 1,0}$ &
  $\mp\frac{\sqrt{6}}{2}\sin\theta\cos\theta$ \\[1mm] \hline}
{\label{dfunc} Explicit forms of $J=2$ $d$ functions.}

\section{Angular distributions for $G\to VV\to(f\bar f)(f\bar f)$}
\label{sec:gravitondecay}

In the heavy graviton mass limit
\big($\beta'=\sqrt{1-4m_V^2/M^2}\to 1$\big), where the only three
amplitudes among the nine $G\to VV$ amplitudes in table~\ref{MGVV} are relevant, 
the differential distribution of the massive-graviton decays, 
$G\to VV\to (f_1\bar f_3)(f_2\bar f_4)$, for the initial polarization
$\lam=\pm 2$ along the $z$-axis is given  by the expression
\begin{align}
 &\hspace{-1.5cm}\frac{d\Gamma_{G}^{\lambda=\pm 2}}
       {d\cos\Theta\,d\cos\theta_1\,d\cos\theta_2\,d\Phi_{12}} \nn\\ \sim
 & (1+6c^2+c^4)\{(1+c_1^2)(1+c_2^2)+4\kappa_1\kappa_2c_1c_2\}
   +2s^4s_1^2s_2^2 \nn\\
 &\mp 8c(1+c^2)\{\kappa_1c_1(1+c_2^2)+\kappa_2(1+c_1^2)c_2\} \nn\\
 &+4s^2s_1s_2\{(1+c^2)(\kappa_1\kappa_2+c_1c_2)\mp 2c(\kappa_2c_1+\kappa_1c_2)\}
      \cos\Phi_{12} \nn\\
 &+s^4s_1^2s_2^2\cos 2\Phi_{12}
\label{Gdecay2}
\end{align} 
with $\Phi_{12}=\phi_1+\phi_2$, the abbreviations $c=\cos\Theta$, 
$s=\sin\Theta$, $c_i=\cos\theta_i$ and $s_i=\sin\theta_i$, and
$\kappa_i$ in eq.~(\ref{eta}); see fig.~\ref{fig:frame} for the
definition of the kinematical variables.
Here, primes ($'$) which indicate the decay variables in the text 
are omitted for simplicity.
For the  $\lam=\pm 1$ state, the distribution reads
\begin{align}
 &\hspace{-1.5cm}\frac{d\Gamma_{G}^{\lambda=\pm 1}}
       {d\cos\Theta\,d\cos\theta_1\,d\cos\theta_2\,d\Phi_{12}} \nn\\ \sim
 & 4(1-c^4)\{(1+c_1^2)(1+c_2^2)+4\kappa_1\kappa_2c_1c_2\} 
         +8s^2c^2s_1^2s_2^2 \nn\\
       &\mp 16c(1-c^2)\{\kappa_1c_1(1+c_2^2)+\kappa_2(1+c_1^2)c_2\} \nn\\
       &-16s^2cs_1s_2[c(\kappa_1\kappa_2+c_1c_2) \mp (\kappa_2c_1+\kappa_1c_2)]
   \cos\Phi_{12} \nn\\
    & -4s^4s_1^2s_2^2 \cos 2\Phi_{12},
\label{Gdecay1}
\end{align}
where the same normalization is used as in eq.~(\ref{Gdecay2}).


\end{document}